\documentclass[fleqn,usenatbib]{mnras}

\usepackage{newtxtext,newtxmath}

\usepackage[T1]{fontenc}

\DeclareRobustCommand{\VAN}[3]{#2}
\let\VANthebibliography\thebibliography
\def\thebibliography{\DeclareRobustCommand{\VAN}[3]{##3}\VANthebibliography}

\usepackage{graphicx}	
\usepackage{amsmath}	
\usepackage{adjustbox}
\usepackage{subcaption}
\usepackage{caption}
\usepackage{multirow}
\usepackage{graphics}
\usepackage{comment}
\usepackage{float}

\newcommand{\msun}{\,M$_{\odot}$}

\newcommand{\mtot}{\,M$_{\mathrm{tot}}$}
\newcommand{\mnsone}{\,M$_{\mathrm{NS1}}$}
\newcommand{\mnstwo}{\,M$_{\mathrm{NS2}}$}
\newcommand{\porb}{\,$P_\mathrm{orb}$}

\title[Heavy DNSs II]{Formation of heavy double neutron stars II: the role of heavy first-born neutron stars and low metallicity}

\author[Nair et al.]{
Ashwathi Nair$^{1,2}$\thanks{E-mail: ashwathinair@swin.edu.au},
Simon Stevenson$^{1,2}$,
\\
$^{1}$OzGrav : The ARC Centre of Excellence for Gravitational Wave Discovery, Hawthorn, VIC 3122, Australia\\
$^{2}$Centre for Astrophysics and Supercomputing, Swinburne University of Technology, Hawthorn, VIC 3122, Australia\\
}

\date{Accepted XXX. Received YYY; in original form ZZZ}

\pubyear{\the\year{}}

\begin{document}
\label{firstpage}
\pagerange{\pageref{firstpage}--\pageref{lastpage}}
\maketitle

\begin{abstract}
The high total mass of GW190425 challenges our understanding of double neutron star (DNS) formation, as no such heavy ($\ge 3$\,M$_\odot$) DNS system has been observed in the Milky Way disk.
Numerous formation scenarios have been proposed to explain its formation.
We test these within a self-consistent binary evolution framework calibrated to the Galactic DNS population.

In Paper~I of this series, we studied the evolution of a $1.4$\,M$_\odot$ neutron star (NS) in a binary with a $2.5$--$10$\,M$_\odot$\ helium star at solar metallicity ($Z = Z_\odot$), assuming Eddington-limited accretion onto the NS using \texttt{MESA}.
In this paper, we consider the evolution of NS-He star binaries with a broad range of NS masses from $1.1$--$1.9$\,M$_\odot$ at $Z=Z_\odot$, 0.1\,$Z_\odot$ and 0.01\,$Z_\odot$.
We find that the formation of heavy DNSs is rare, accounting for only $\sim 0.5$ per cent of DNSs.
The majority of these are formed through the `standard formation' channel, widely believed to form GW170817 and observed Galactic DNSs. We find no contribution through the `fast-merger' channel at solar metallicity, but systems at low metallicity predominantly form through unstable mass transfer. Furthermore, we find no significant dependence of the heavy DNS formation fraction on metallicity over the range considered here. We conclude that heavy DNSs do not form a separate subpopulation and merely represent the high-mass tail of the standard DNS population.
\end{abstract}




\section{Introduction}
\label{sec:intro}

The Fourth LIGO \citep[][]{LIGO:2015CQG}, Virgo \citep[][]{2015CQGra..32b4001A}, and KAGRA \citep[][]{KAGRA:2019NatAst} observing run (LVK O4) lasted from May 2023 to November 2025. 
The LVK reported approximately 128 and 161 significant merger events from the O4a and O4b run, respectively, doubling the number of detections from the previous three runs \citep[][]{Abbott:2023PRXGWTC3,LIGOScientific:2025slb, LVK2026GWTC5Intro, LVK2026GWTC5Results}.

The vast majority of new gravitational-wave (GW) events reported in the Fourth and Fifth Gravitational-Wave Transient Catalogs  \citep[GWTC-4.0 and GWTC-5.0;][]{LIGOScientific:2025slb, LVK2026GWTC5Intro, LVK2026GWTC5Results} are binary black holes \citep[BBHs; e.g.,][]{LIGOScientific:2025rid, LIGOScientific:2025brd, LIGOScientific:2025rsn}, with one black hole-neutron star (BH-NS) binary \citep[][]{LIGOScientific:2024elc}. 
No significant detections of double neutron star (DNS) mergers were made during the O4 run. 
Recently, \citet{Niu:2025arXiv} reported a sub-threshold candidate, GW231109\textunderscore235456, as a potential astrophysical DNS merger using a DNS-targeted search, in contrast to the GWTC-4.0 analyses, which look for all three types of compact binary mergers \citep[][]{LVK2026GWTC5Methods}. 
This leaves us with only two significant DNS candidates, GW170817 \citep{Abbott:2017PhRvL} and GW190425, which was detected in the third observing run \citep[O3;][]{Abbott:2020ApJL}, with no change in the DNS population detected via GWs. 

DNSs residing in our Milky Way galaxy have also been detected using pulsar timing \citep[][]{HulseTaylor:1975ApJL}. 
As of today, there are 36 DNS systems \citep[][]{NairStevenson:2025MNRAS}, including two newly discovered systems, PSR J0641+0448, with a total mass (\mtot{} = \mnsone{} + \mnstwo{}) of $2.58^{+0.029}_{-0.031}$\msun{}\,\citep{Yang_2026} and PSR J1856--0039, with a \mtot{} of 2.48\,\msun{} \citep[][]{Yang:2026arXivJ1856}.
The DNS population observed in our Galactic disk and the GW population continue to dominate the narrow total mass range of $2.3$--$2.9$\,\msun{}
except GW190425, which has total mass of $3.4^{+0.3}_{-0.1}$\msun{}. We have not observed such heavy DNS systems (\mtot{} $\geq$ 3\,\msun{}) in the Galactic disk, although they have been observed in globular clusters \citep[GCs; e.g.,][]{Padmanabh_2024, Barr_2024}. However, theoretical models predict that dynamical formation of DNSs is inefficient in star clusters and cannot contribute significantly to the observed GW population \citep{Ye:2020ApJL}.

The majority of DNS are believed to form through the isolated evolution of massive binary stars. The commonly accepted isolated DNS formation pathway has existed for decades \citep[e.g.,][]{1975A&A....39...61F, 1991PhR...203....1B, 2023pbse.book.....T}. In this scenario, two hydrogen-rich zero-age main-sequence (ZAMS) stars start their evolutionary journey together. The initially more massive star, the primary, transfers mass onto the companion star and undergoes a supernova (SN) explosion, leaving behind a first-born neutron star (NS1). The secondary eventually expands and fills its Roche lobe and begins transferring mass to the companion NS. As a result of the large mass ratio, the mass transfer is unstable and leads to common envelope evolution  \citep[CEE;][]{1976IAUS...73...75P, 1993PASP..105.1373I, 2013Sci...339..433I}. 
CE is thought to be the dominant channel for DNS formation \citep[see][]{vanSon:2025ApJ, broekgaarden2026commoncommonenvelopesquantifying}. If the binary successfully ejects the hydrogen envelope, it leaves a helium-rich star and NS1 behind.  
 Depending on its mass, the helium star (He star) may reexpand and initiate another phase of mass transfer while burning helium in its shell, namely Case BB Roche lobe overflow (RLO). This process strips the helium envelope and leaves behind a pre-SN metal core. The metal core then undergoes an ultra-stripped supernova (USSN) and leaves behind a second neutron star (NS2), giving rise to a DNS merger. 
This formation pathway is what we refer to as the `standard formation pathway' throughout this article. Systems believed to form via the standard formation pathway include the majority of the observed Galactic population of DNSs \citep[][]{Tauris:2017ApJ} and GW170817 \citep{Abbott:2017PhRvL}. 

However, the formation of a heavy DNS system such as GW190425 remains uncertain. No DNS systems with a total mass of \mtot{} $\geq$ 3\,\msun{} have been observed in the Galactic field. Any model proposed to explain the formation of GW190425 must explain both the formation of merging heavy DNS systems and the lack of observation of heavy DNSs in the Galactic field. 

Whether such massive DNSs can form through isolated binary evolution at solar metallicity has been discussed \citep[e.g.,][]{Kruckow:2020A&A, Galaudage_2021, Mandel:2021MNRAS}. 
One proposed explanation is that these heavy DNS systems might form through a `fast-merger' channel \citep[][]{Abbott:2020ApJL}. In this scenario, the progenitor binary that survived the first CE phase undergoes a second unstable mass transfer leading to another CE phase \citep{2003ApJ...592..475I, Belczynski_2002, Vigna-Gomez:2018MNRAS}, producing an extremely compact DNS with an orbital period of less than an hour. Such systems merge rapidly through GW emission and are difficult to detect through radio pulsar timing surveys because of their short lifetimes and the strong acceleration effects that reduce the sensitivity of pulsar surveys. 
Several studies have explored whether this fast-merger channel can account for the unusually high total mass of GW190425. For example, \citet{Abbott:2020ApJL} discuss the possibility of GW190425 being a fast-merging DNS \citep[see also:][]{RomeroShaw:2020MNRAS, Safarzadeh:2020efa}. 

\citet{VignaGomez:2021ApJL} introduced an alternate formation channel in which a massive ($M_\mathrm{He} \geq 9$\,\msun{}) He star produces a heavy NS2. Due to its limited radial expansion, the heavy helium star does not transfer mass onto its companion neutron star, avoiding recycling the pulsar. After a few tens of millions of yr, non-recycled pulsars become radio-quiet, such that these binary systems can only be detected by GW observatories, which hints at why such heavy DNSs might not be observable at radio wavelengths.

\citet{Qin:2024A&A} proposed the formation of massive GW190425-like DNSs through stable case BB/BC\footnote{Case BC MT is when a He-rich star transfers mass onto the companion when carbon is ignited in the stellar core} MT from a He-rich star onto the companion NS, requiring a massive first-born NS. Observational evidence suggests that some NSs can be born massive depending on their final pre-supernova metal core mass \citep[][]{Valentim_2011, Kiziltan_2013, antoniadis2016millisecondpulsarmassdistribution, 2016MNRAS.460..742M, 2018ApJ...860...93S} and as inferred from several binary pulsars such as PSR J1640$+$2224 \citep{2020ApJ...892....4D} and PSR J1614-2230 \citep{2011MNRAS.416.2130T}.
This motivates treating the birth mass of the first-born neutron star as a free parameter, particularly when exploring formation channels that can produce high-mass DNS systems. \citet{Zhang:2023MNRAS} proposed super-Eddington accretion, as high as 1000$\dot{M}_\mathrm{Edd}$ during a Case BB RLO mass transfer onto the companion NS as a possible scenario to form  GW190425-like DNSs.

In addition to binary-interaction physics, metallicity plays a crucial role in shaping the properties of compact remnants. One of the important metallicity-dependent effects in massive stars is mass loss through radiation (line)-driven winds \citep[e.g.,][]{1992ApJ...401..596L,  2001A&A...369..574V, Vink_2021}. At lower metallicities, weaker winds lead to reduced mass loss, allowing stars to retain more mass prior to core collapse and form a more massive remnant. 
Due to the long delay times between their formation and merger, merging double compact objects are formed at different times with varying metallicities in our Universe. 
Consistent with this, \citet{Giacobbo:2018MNRAS} showed that massive DNSs with total masses in the range $3.2$--$3.5$\,\msun{} tend to form at lower metallicities (Z $<$ 0.008) through the isolated formation channel. While the formation and detection efficiency of BBH are strongly metallicity dependent \citep[][]{2010ApJ...715L.138B, Giacobbo:2018MNRAS,Mapelli:2018MNRAS,Santoliquido:2021MNRAS,Broekgaarden:2022MNRAS}, the formation efficiency of DNSs is typically found to be weakly dependent on metallicity, although significant uncertainties remain \citep[][]{Giacobbo:2018MNRAS, Gallegos-Garcia:2023ApJ, vanSon:2025ApJ}. 

In the first paper of this series \citep[][hereafter Paper I]{NairStevenson:2025MNRAS}, we studied the evolution of He star + NS binaries that formed DNSs with Eddington-limited accretion of He-rich material onto the NS1 at solar metallicity, and investigated the progenitors of GW190425-like systems. We assumed a fixed NS1 mass of 1.4\,\msun{} and the mass of NS2 was determined from the mass of its progenitor. One of the aims was to ensure that our models could explain the population of observed Galactic DNS formed in the Galactic disk. 
We found that within this restricted parameter space, we could not form enough heavy DNSs to explain the observation of GW190425-like systems. Nevertheless, among the massive DNS systems that did form, the standard formation channel was dominant, while formation via unstable Case BB mass transfer was disfavoured.

The limited parameter space was not enough to test all the uncertainties associated with the formation of GW190425. Therefore, in this work we extend our previous study by broadening the parameter space to include variations in NS1 birth mass and metallicity. This allows us to self-consistently test the different formation channels and identify which of the formation channels are more likely to form heavy GW190425-like systems. Variation to allow super-Eddington accretion is out of scope for this paper.

We arrange the rest of the paper as follows:
In Sec.~\ref{sec:methods}, we describe our modelling framework and the initial parameter space. In Sec.~\ref{sec:results}, we present results on how our models at solar metallicity compare to the observed DNS population when we combine the effects from the extended parameter space. We discuss the
GW190425 formation pathways at solar and low metallicities in detail and determine the formation channel that is more likely to form heavy GW190425-like DNS. In Sec.~\ref{sec:discussion} we discuss our results and potential caveats and uncertainties in detail by comparing our results with other studies. 
Finally, we conclude our analyses and findings in Sec.~\ref{sec:conclusions}.

\section{Methods}
\label{sec:methods}

\subsection{He Star-NS Binary Evolution}
\label{sec:model_calculation}

\subsubsection{Detailed Evolutionary Models}
\label{subsubsec:detailed_models}

Unstable mass transfer initiated by the initially less massive star (secondary hydrogen-rich star) gives rise to a hydrogen common envelope and engulfs the companion star. If the binary successfully ejects the hydrogen envelope, upon survival, it leaves behind a binary consisting of a stripped helium-rich star and a companion first-born neutron star (NS1). In this study, we assume that the binaries have survived the CE phase and consider the post-CE phase, i.e., the helium star $-$ neutron star (He-star + NS) binary as our starting point of evolution. 

In order to test the mentioned formation channels we perform detailed binary-star evolution using the one-dimensional (1D) stellar evolution code \texttt{MESA} \citep[][]{Paxton2011,Paxton2013,Paxton2015,Paxton2018,Paxton2019,Jermyn2023}, version \texttt{23.05.01}.

\begin{figure}
    \centering
    \includegraphics[width=\columnwidth]{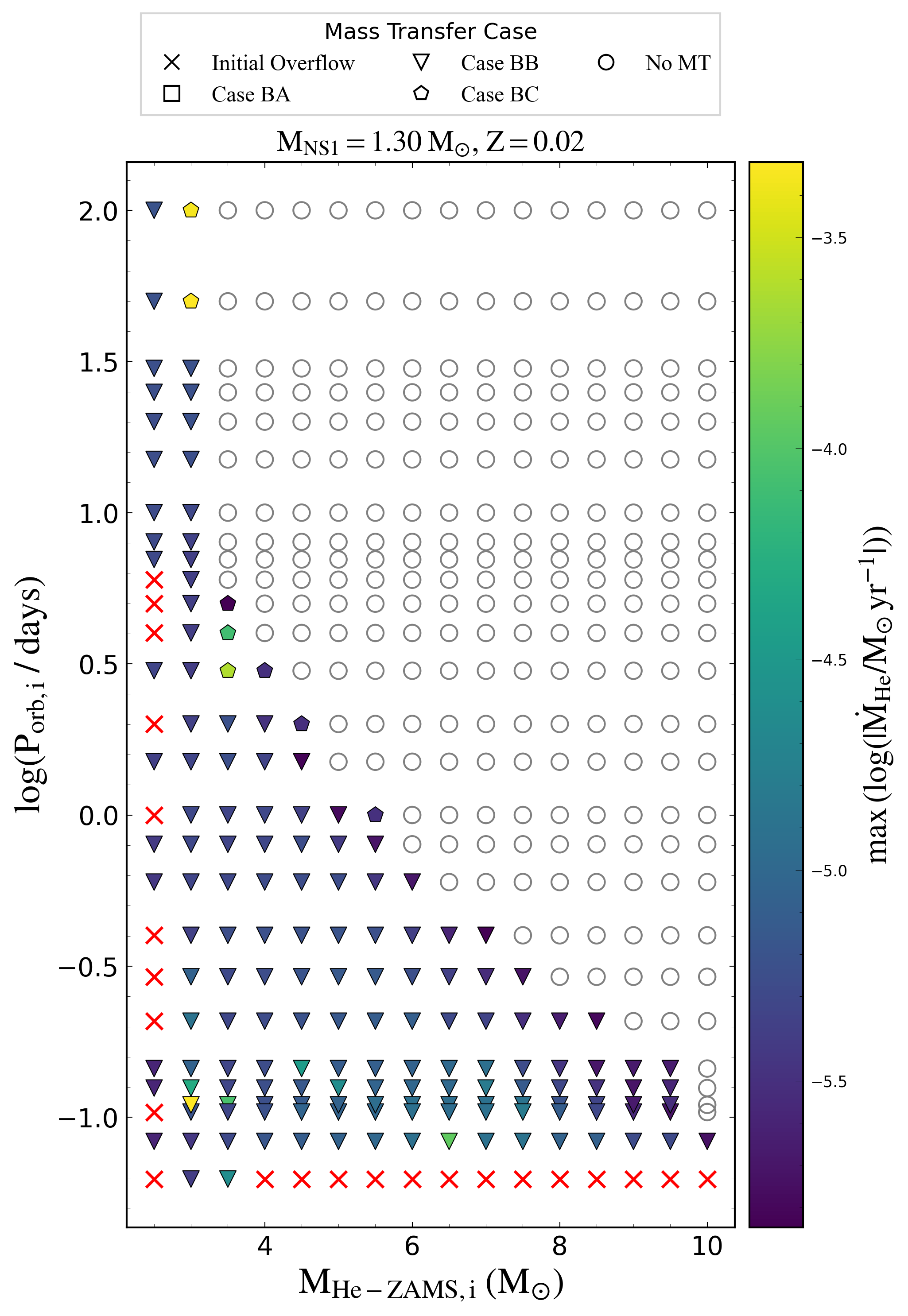}
    \caption{Parameter space of initial orbital period (log scale) as a function of initial He-star mass, for a companion neutron star mass of 1.30\,\msun{} and metallicity Z = 0.02. Different symbols indicate different phases of mass transfer: grid points on the parameter space with red solid crosses represent where the He star already overfills its Roche lobe at the start of evolution (Initial Overflow); squares, inverted triangles and pentagons correspond to systems undergoing Case BA, BB, or BC mass transfer, respectively; open circles correspond to systems that do not undergo mass transfer. The colorbar on the right shows the peak mass transfer rate (log scale) from the He-star donors for systems that do undergo mass transfer, gradually increasing from purple to yellow.}
    \label{fig:porb_mhe_mtrate}
\end{figure}

The assumptions we use to model He stars and He star + NS binaries follow \citet{NairStevenson:2025MNRAS} except where noted in this section.
We model the He-rich stars with a helium mass fraction $Y= 1 - Z$, at metallicity of $Z = Z_\odot = 0.02$ (solar metallicity), 0.1\,$\mathrm{Z_{\odot}}$, and 0.01\,$\mathrm{Z_{\odot}}$, to simulate the stars that might have formed at different times and different environments in the universe. 
\citet{gotberg2023stellarpropertiesobservedstars} report the hydrogen mass fraction on the recently observed stripped helium stars in the mass range of $2$--$8$\,\msun{} \citep{Drout:2023Sci}, hinting that a small amount of hydrogen can be retained after the CE phase. Additionally, \citet{Nie:2025ApJ} have also shown that a small amount of hydrogen ($\leq 1$\,\msun) may be retained on the surface of the helium star.
The He-star models were initialized with masses ranging from $2.5$--$10$\,\msun{}, with $\Delta M_\mathrm{{He,\,i}} = 0.5$. He-stars were evolved from pre-main sequence (pre-MS) to the He-ZAMS using the \texttt{MESA} module \texttt{MESAstar}. During this evolution, mass loss via stellar winds was active. As a result, at solar metallicity, the He-star models ($> 3$\,\msun{}) undergo slight mass loss before reaching the He-ZAMS phase, and thus the final He-ZAMS masses span $2.5$--$9.8$\,\msun{}. Low-metallicity He-stars do not lose a significant amount of mass during the pre-MS, thereby maintaining the mass range from $2.5$--$10$\,\msun{}. We modelled our He-star models to evolve until central carbon depletion.
We model our He-star in a binary with the first-born neutron star, with masses ranging from 1.1 to 1.9\,\msun{}  using the \texttt{MESA} module \texttt{MESAbinary}. The He star + NS binaries cover a grid of parameter space of orbital period ($P_\mathrm{orb}$) ranging between 0.06 d ($\sim$ 1.4 hours) and 100 d (see Fig~\ref{fig:porb_mhe_mtrate}). We assume that the mass accretion rate for He-rich material onto a neutron star is Eddington-limited, $\dot{M}_\mathrm{Edd} = 3 \times 10^{-8}$\,M$_{\odot}$\,yr$^{-1}$.

Modelling the helium star wind mass-loss rate is challenging due to the observational scarcity of stripped helium stars.
Within MESA, the standard Dutch wind mass loss recipe for hot stars with effective temperature ($\mathrm{T_{eff}}$) > $10^4$ K uses mass loss rates given by \citet{2000A&A...360..227N}. We use the default stellar wind loss recipe by \citet{2000A&A...360..227N} to model the mass loss of He stars due to winds at solar metallicity. The mass-loss rates strongly depend on the luminosity and the chemical composition of the star and are given as
$\mathrm{\dot{M}}\,\propto\,\mathrm{Z}^{\beta}$, 
where $\beta$ = 0.46. 
However, in practice, models implementing the \citet{2000A&A...360..227N} mass-loss rates have a weak metallicity scaling due to the self-enrichment process (and the dependence on the current value of $Z$ rather than the initial $Z$) and the mass loss due to wind is overestimated. \citet{Sander_Vink_Higgins_Shenar_Hamann_Todt_2020} tested different wind mass loss prescriptions at solar metallicity and did not find any major difference between them. However, at low metallicity, the models with \citet{2000A&A...360..227N} overestimated mass loss rates. 
In order to take this into account when we model helium stars at low metallicity, we choose to model helium star mass-loss rates (implemented using the \texttt{other\textunderscore hooks}) using Eq. 9 of \citet{2010ApJ...714.1217B} with the metallicity dependence given by \citet{2005A&A...442..587V} where $\beta = 0.86$. See Sec.~\ref{subsec:discussion_b} for a detailed discussion on He star wind mass-loss rates.

Depending on the He star mass, the companion NS mass, and their orbital separation, the binary undergoes different mass transfer processes. In Fig.~\ref{fig:porb_mhe_mtrate}, we show an example of our simulated grid of He star--NS binaries for an NS1 companion of 1.3\,\msun{} at solar metallicity in a \porb{}--$M_{\rm{He,i}}$ parameter space. Low-mass helium stars ($\sim$\,2.5\,\msun{}--3\,\msun{}) experience significant radial expansion during their evolution (see Fig.2 from \citealt{NairStevenson:2025MNRAS}), allowing them to initiate mass transfer for a broad range of initial orbital periods. As the initial He star mass increases, the evolutionary time-scale becomes shorter, resulting in a short-lived mass transfer phase. Many systems remain detached, particularly at wider orbital separations,  because the helium star does not expand sufficiently to fill its Roche lobe prior to the SN explosion. Most of the systems in our simulation undergo Case BB RLO. 
Only a small fraction of systems at the boundary between systems that undergo Case BB RLO mass transfer and no mass transfer initiate mass transfer when the star is burning carbon in its core, namely Case BC RLO mass transfer.

We set a criterion for identifying if the systems undergo unstable mass transfer depending on the peak mass transfer rate from the He star, where, if the $\max\,(\log (|\dot{M}_\mathrm{He}$/\msun{}\,$\mathrm{yr}^{-1}|))$ $>$ -1, we assume that the system is undergoing an unstable mass transfer (similar condition has been adopted in \citealt{Fragos:2022jik}). 
Based on this, we find that all  systems undergo stable mass transfer at solar metallicity. This is true for our He stars in binary with high-mass NS1 models as well. However, the stability of mass transfer varies for binaries at low metallicity ($Z = 0.1$\,$\mathrm{Z_{\odot}}$ and 0.01\,$\mathrm{Z_{\odot}}$). 

As the He-star transfers mass to its companion NS, its He envelope is stripped. The final He-envelope mass depends on the initial mass, the evolutionary stage at which mass transfer begins, and the duration of the mass transfer episode (refer to sec. 2.3 and Fig. 6 of \citealt{NairStevenson:2025MNRAS} for a detailed explanation). The mass transfer phase in low-mass He stars lasts longer, as they expand more, thereby stripping away more He envelope and leaving behind a low-mass carbon-oxygen metal core. 

\subsubsection{Post-SN orbital dynamics}
\label{subsubsec:postSNorbit}

Depending on the final CO core mass and composition of the He star, the star undergoes an SN explosion and leaves behind a compact remnant. The nature of the SN explosion (including the mass ejected and the natal kick velocity experienced by the newborn neutron star) determines whether the binary remains bound or gets disrupted. 
To calculate the remnant mass of the newborn NS (NS2) and the natal kick velocity, we use the `Modified model' introduced in \citet{NairStevenson:2025MNRAS}, where we modified the analytical prescription given by \citet{Mandel:2020MNRAS}. These modifications were introduced to explain the observed Galactic DNS population using our simulated DNSs. We incorporated the updated conditions for the USSN and ECSN, along with the updated calibration on the kick prescription given by \citet{Kapil:2023MNRAS}. For detailed information on the modifications made to the \citet{Mandel:2020MNRAS} prescription, we direct the readers to sec. 3.1.2 of \citet{NairStevenson:2025MNRAS}. 

We assume that the pre-SN CO core metal in a binary with NS1 explodes instantaneously \citep{Brandt1994TheEO}. This SN explosion significantly alters the orbital dynamics of the binary as a result of instantaneous mass loss and the natal kick velocity experienced by the NS2. We assume an isotropic distribution of kick directions. Using Eq. (2.1) -- (2.8) from \citet{Brandt1994TheEO} we determine whether the binary system remains bound or not, and calculate the post-SN orbital dynamics of the surviving DNS systems.

\subsection{Initial simulation grid and distribution functions}
\label{sec:initial_grid}

To infer the population properties of DNS systems, we apply the aforementioned post-processing framework to reweight our population according to the chosen distributions of He star mass, NS1 mass, and orbital period. For a detailed explanation of how this is performed, we point our readers to Sec. 3.1 of \citet{NairStevenson:2025MNRAS}. Here, we describe the initial distribution models we consider while performing population synthesis.

\textit{Helium star distribution:} For the distribution of initial He star masses, we assume a `power law' distribution $p(M_\mathrm{He}) \sim M_\mathrm{He}^{-\alpha}$, inspired by the stellar initial mass function \citep[IMF;][]{1955ApJ...121..161S}, where $\alpha = 2.35$.

\textit{Orbital period distribution:} We assume a simple model `flat in the log' distribution $(P(P_\mathrm{orb}) \sim 1/P_\mathrm{orb})$ for the initial orbital period distribution.

\textit{NS mass distribution:}

\begin{figure}
    \centering
    \includegraphics[width=\columnwidth]{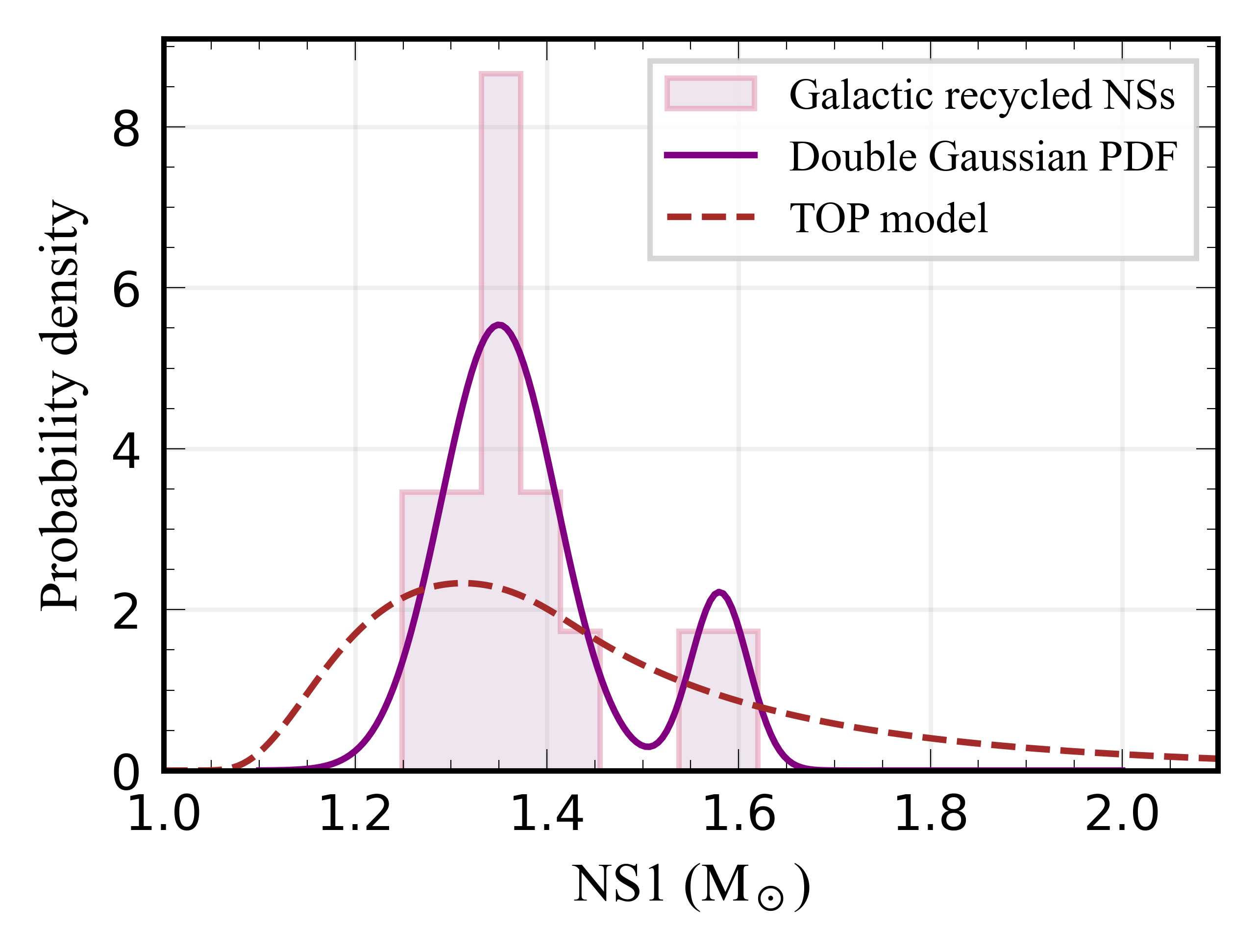}
    \caption{Mass distribution of the first-born neutron star (NS1). The shaded histogram shows the observed sample of recycled neutron stars in double neutron star systems observed in the Galactic field. The solid purple curve corresponds to a double-Gaussian probability density function fit to the observed masses, peaking at the two known sub-populations near $1.35$\,\msun{} and $1.6$\,\msun{}. 
    The dashed brown unimodal distribution curve shows the predicted NS1 mass distribution fitted with the `turn-on' power law (ToP) model introduced in \citet{YouZhu:2025NatAst} based on the heterogeneous observed sample of $\sim 90$ precisely measured NSs in binaries.}
    \label{fig:MNS1_distribution}
\end{figure}

Various uncertainties and poorly constrained physical processes in DNS evolution, such as mass transfer, supernova physics, natal kick velocities, and the neutron star equation of state, lead to significant uncertainty in the predicted neutron star mass distribution in DNS systems.
Observationally, the study of the mass distribution of NSs in DNS binaries has been conducted for more than a decade. 
First, with a limited sample of nine DNS binaries by \citet{Ozel:2012ax} and \citet{Kiziltan_2013}, the distribution was explained by a single-component Gaussian fit. Using an updated sample size of 17 DNSs observed in our Galaxy, \citet{Farrow:2019xnc} inferred that the distribution represents a bimodal nature. With the current sample size of $\sim$26 confirmed DNSs in our Galactic field, and $\sim$14 precisely measured recycled pulsars, the recycled neutron stars in our Galactic DNS binaries range from $1.2$ to $\sim$1.62\,\msun{}.  We fit the distribution of recycled NS masses with a double-Gaussian (DG) model, which is shown in Fig.~\ref{fig:MNS1_distribution}.

Additionally, NSs are born in different environments and can be observed in a wide range of binaries at different stages of binary evolution. Studies by \citet{valentimMassDistributionNeutron2011a, Kiziltan_2013, antoniadis2016millisecondpulsarmassdistribution, Alsing:2017bbc} have looked at the distribution of NSs in such binaries, pointing towards a bimodal distribution with a first narrow peak at $\sim$1.3\,\msun{} and a second broader peak at $\sim$1.8\,\msun{}.  
Recently, using the mass measurements of 90 NSs observed in various binaries, \citet{YouZhu:2025NatAst} showed that a unimodal turn-on power law (ToP) model fit can explain the distribution of the observed NS population (also shown in Fig.~\ref{fig:MNS1_distribution}).

These weights are incorporated into the population synthesis framework when constructing the final distributions of total masses and orbital properties, ensuring that the simulated population reflects the distribution of NS1 masses.

\section{Results}
\label{sec:results}

\subsection{Orbital period -- eccentricity distribution at solar metallicity}
\label{subsec:Porb_e_sol}

\begin{figure}
    \centering
    \includegraphics[width=\columnwidth]{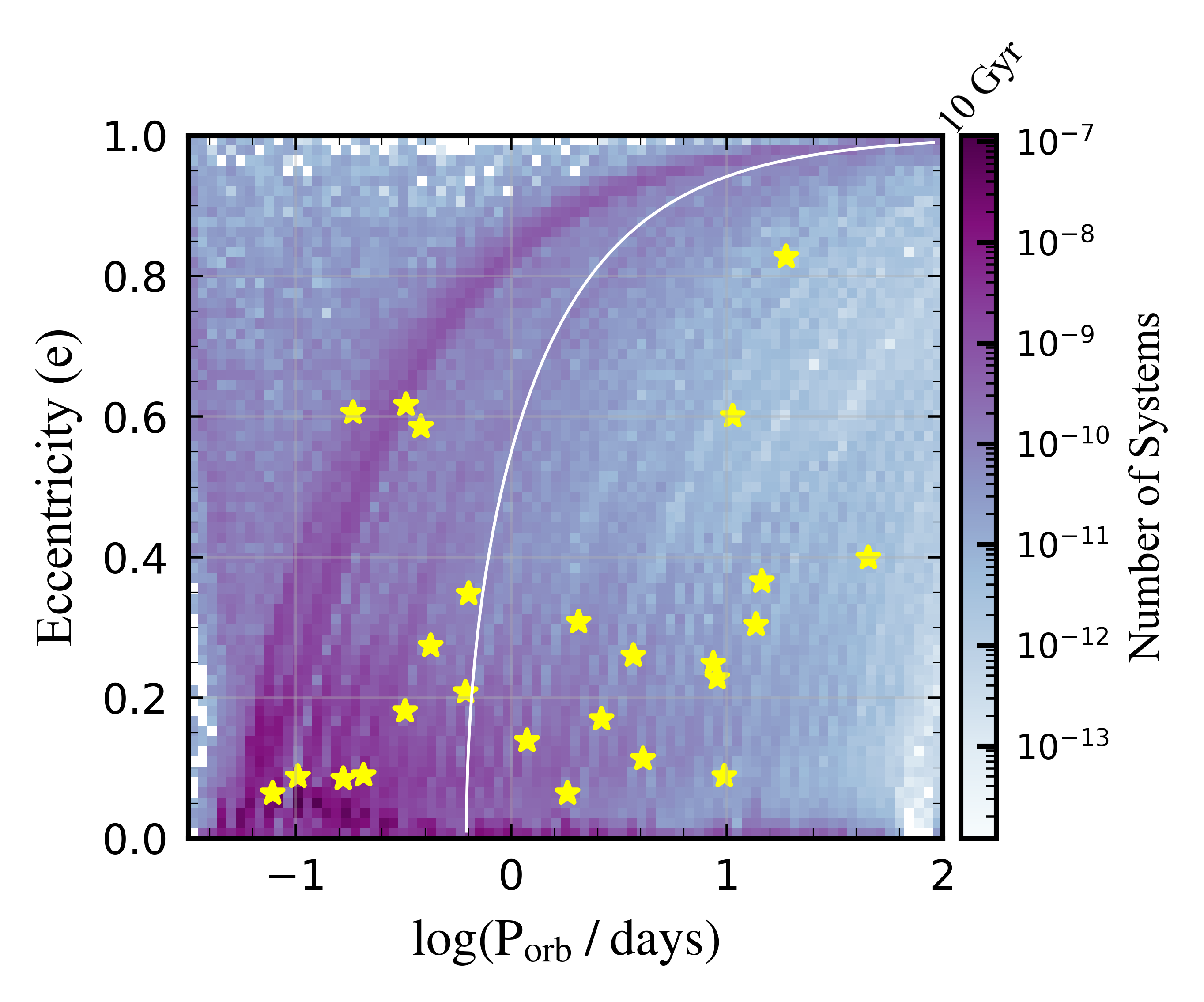}
    \caption{The combined weighted 2D distribution of post-supernova eccentricity (e) as a function of orbital period $P_{\rm{orb}}$ for $\sim$ 6000 simulated double neutron stars with varying first-born neutron stars in a binary with helium stars at solar metallicity compared with the observed double neutron star systems in our Galactic field (yellow stars). The white solid line corresponds to a constant merger time of 10 Gyr.}
    \label{fig:interpolated_porb_e}
\end{figure}

Fig.~\ref{fig:interpolated_porb_e} shows the combined post-SN birth distribution of the system's orbital period and eccentricity (\porb{}--$e$) of all simulated systems at solar metallicity, with the contribution from NS1 with varying masses.
In fig.13 from \citet{NairStevenson:2025MNRAS}, we showed how the post-SN orbital dynamics in a \porb{}--e distribution broadly matched the observed DNS population with just an NS1 mass of 1.4\,\msun{}. Most of the observed systems can be reproduced by He stars of masses $3$--$3.5$\,$\mathrm{M_{\odot}}$ that undergo electron-capture supernova (ECSN) and ultra-stripped supernova (USSN). The newborn NS receives a low kick of $\sim$20--$50$\,km\,s$^{-1}$, and all DNS systems with short (few hours) and long orbital periods ($\sim$100\,d) stay bound. The three systems PSR\,J1757$-$1854, PSR\,J0509$+$3801, and PSR\,B1534$+$12, observed in the short orbital period and high eccentricity parameter space, however, are not produced by the above mechanism. They are observed to be formed by our He stars of masses $>$ 4\,$\mathrm{M_{\odot}}$ and with kick velocity exceeding $100$\,km\,s$^{-1}$. 
This core result stays the same when we combine the contribution from NS1 with varying masses.

\subsection{DNS total mass distribution at solar metallicity}

Intermediate masses not explicitly simulated in the grid due to detailed models being computationally expensive were interpolated using the 1D linear piecewise interpolation method (see Fig.~\ref{fig:interpolated_cdfs} in Appendix~\ref{sec:ns1_interpolation}). This approach allows us to incorporate a complete range of neutron star masses, which more accurately reflects the diversity of first-born neutron stars expected in nature, thereby making the distribution smooth. 

\begin{figure}
    \centering
    \includegraphics[width=\columnwidth]{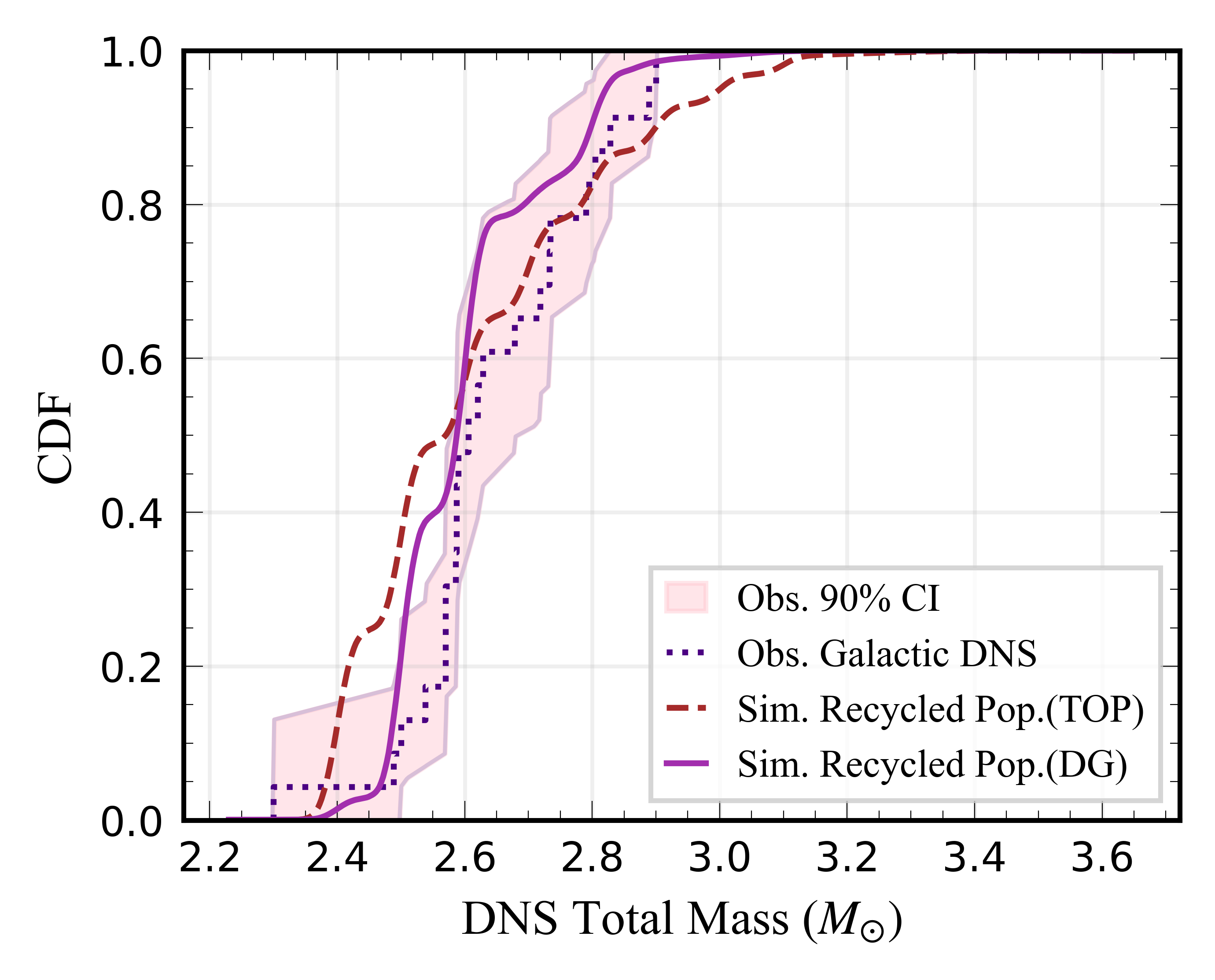}
    \caption{The cumulative distribution function (CDF) of the simulated recycled DNS population at solar metallicity broadly agrees with the observed Galactic DNS distribution (dotted purple). 
    The model weighted using the double-Gaussian NS1 distribution (DG; solid purple) shows the closest agreement, lying largely within the 90 percent observational confidence interval (shaded pink), while the TOP-weighted model (dashed brown) exhibits a slight systematic offset toward lower masses.}
    \label{fig:top_dg_dns}
\end{figure}

Fig~\ref{fig:top_dg_dns} shows the combined CDF for all simulated recycled DNS total mass distributions compared against the observed Galactic DNS total mass distribution. The DNS total mass distribution closely matches that of the observed Galactic DNS population, largely lying within the 90 per cent confidence interval. We overproduce DNSs with total masses of $\sim$2.55\,\msun{}. This is mainly because each simulated CDF is weighted by the initial He star mass and orbital period, where the He star weights are drawn as mentioned in Sec.~\ref{sec:initial_grid}, which favours lower-mass helium stars. These lower-mass helium stars produce NS2 masses with a narrow peak at $\sim$1.2\,\msun{}. The CDFs are then combined using NS1 weights derived from a double Gaussian fit to the observed recycled NS mass distribution in the Galactic field, which peaks at $\sim$1.35\,\msun{}. The combination of these two overproduces DNSs with total mass $\sim$2.55\,\msun{}.

\subsection{GW190425 progenitors}
\label{subsec:GW190425_prog}

In considering a broader parameter space, the resulting DNS population remains broadly consistent with observations. We now compare and identify which of the proposed formation channels for heavy DNSs is more likely to give birth to a GW190425-like heavy DNS (note that heavy DNSs and GW190425-like heavy DNSs are defined as DNSs with $M_{\rm{tot}}$\,$\ge 3.0$\,\msun{}). We calculate the fraction of massive DNSs from each proposed channel. We define the channels for heavy GW190425-like systems based on the initial mass assumption of He star and NS1 in different studies, along with the second mass transfer phase the binary experiences after surviving the CE phase:

\begin{itemize}
    \item Standard formation pathway \citep[][]{Tauris:2017ApJ}: Models with first-born neutron star masses $M_{\rm NS1} < 1.7$\,\msun{} (see Sec.~\ref{sec:intro}) that undergo stable Case BB Roche-lobe overflow mass transfer, survive the second supernova explosion, and successfully form DNS systems.

    \item Massive He-star pathway \citep[][]{VignaGomez:2021ApJL}: Models with initial helium-star masses $M_{\rm He} \geq 8.8$\,\msun{} and varying NS1 masses. We identify systems that avoid further mass transfer prior to core collapse and subsequently form massive DNS systems following the second supernova explosion.

    \item Massive NS1 pathway \citep[][]{Qin:2024A&A}: Models with initially massive first-born neutron stars, $M_{\rm NS1} \geq$ 1.7\,\msun{}, that undergo stable Case BB/BC Roche-lobe overflow mass transfer and successfully form massive DNS systems.

    \item Low-metallicity binaries \citep[][]{Giacobbo:2018MNRAS}: Models of He-star--NS binaries at sub-solar metallicities ($Z = 0.1$\,$\mathrm{Z_{\odot}}$ and 0.01\,$\mathrm{Z_{\odot}}$) that form massive DNS systems.

\end{itemize}

\begin{figure}
    \centering
    \includegraphics[width=\columnwidth]{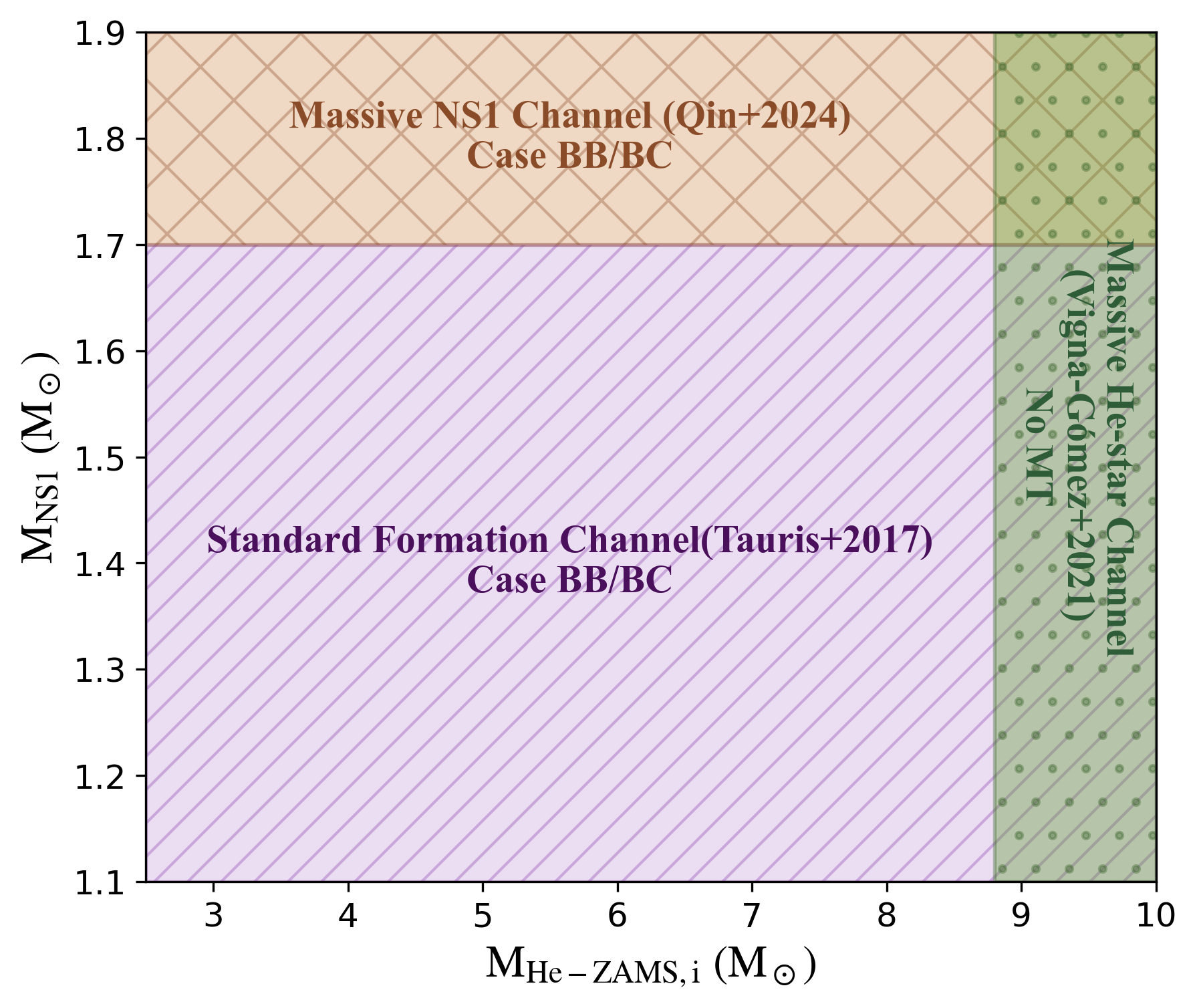}
    \caption{$M_{\rm{He-ZAMS,i}}$--$M_{\rm{NS1}}$ parameter space, showing the regions associated with each of the different proposed formation channels for  DNSs. These are defined based on both the initial masses and the type of mass transfer process that the binary experiences. }
    \label{fig:parameter_space_plot}
\end{figure}

Note that the parametrized definition is purely to calculate the contribution from each individual channel and does not represent a hard boundary for the formation of heavy DNSs in nature.
We show the parameter space for each formation channel in Fig.~\ref{fig:parameter_space_plot}.

\subsubsection{At solar-metallicity}
\label{subsubsec:solar190425}

\begin{figure*}
    \centering
    \includegraphics[width =\textwidth]{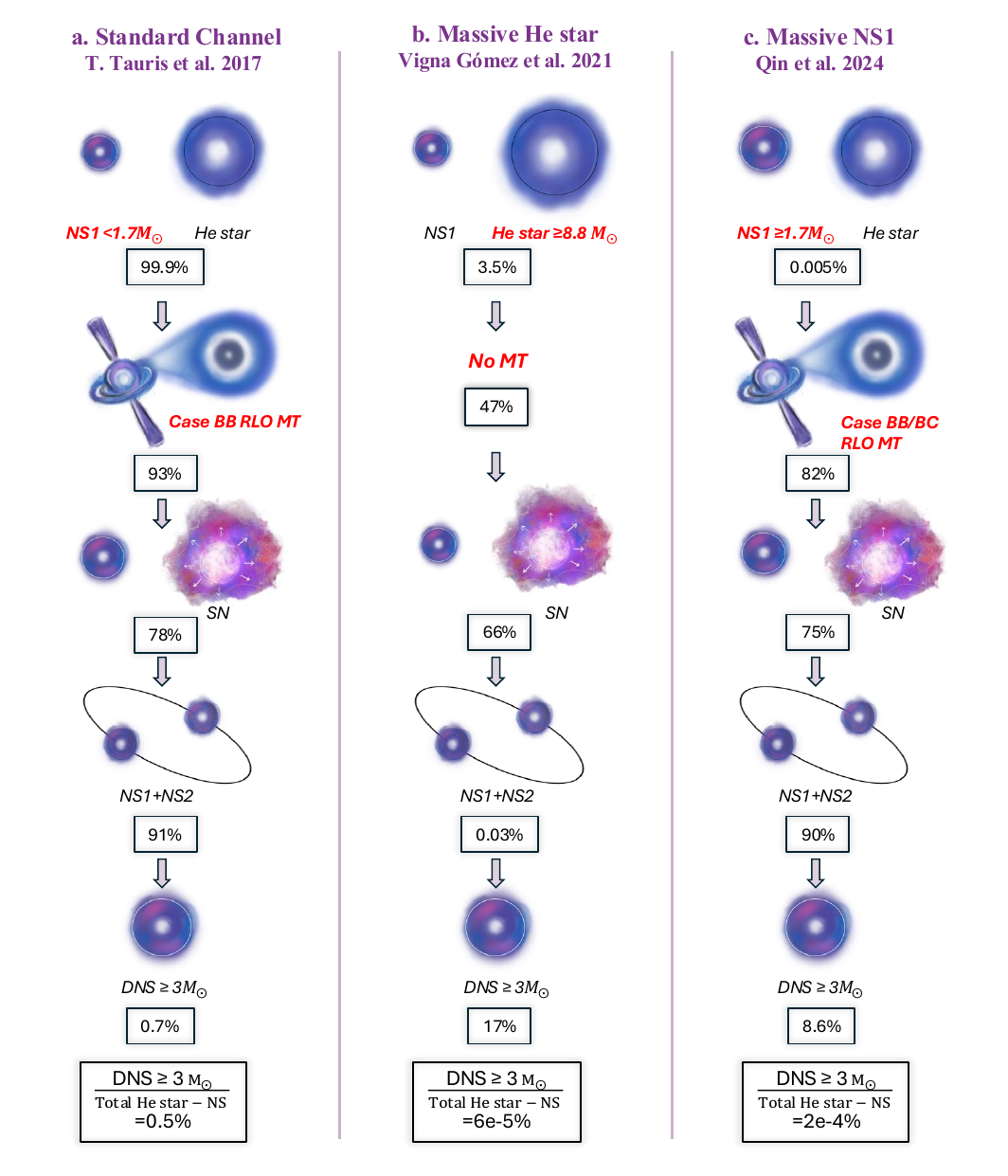}
    \caption{Formation history of a massive double neutron star for three evolutionary pathways (left to right) named as: a. Standard formation pathway \citep{Tauris:2017ApJ}, b. Massive He-star pathway \citep{VignaGomez:2021ApJL}, 3. Massive NS1 pathway \citep{Qin:2024A&A} as defined in Sec.~\ref{subsec:GW190425_prog}. The numbers pointed out using the arrows represent the percentage of simulated binaries that end up in that particular phase out of systems that evolve through the prior phase. For example, out of all simulated systems, 99.9 per cent of them contribute to the standard channel; out of those 99.9 per cent of He star + NS binaries, 93 per cent of all simulated binaries experience stable case BB Roche lobe overflow mass transfer from the He-ZAMS stars onto the companion first-born NS; among those 93 per cent, 78 per cent of systems undergo supernova explosion, out those 78 per cent 11 percent of the systems become double neutron stars, and out of those 11 percent only 0.7 per cent are heavy ($\ge$ 3\,\msun{}) double neutron star mergers.}
    \label{fig:formation_channels_frac}
\end{figure*}
 
\begin{figure}
    \centering
    \includegraphics[width=\columnwidth]{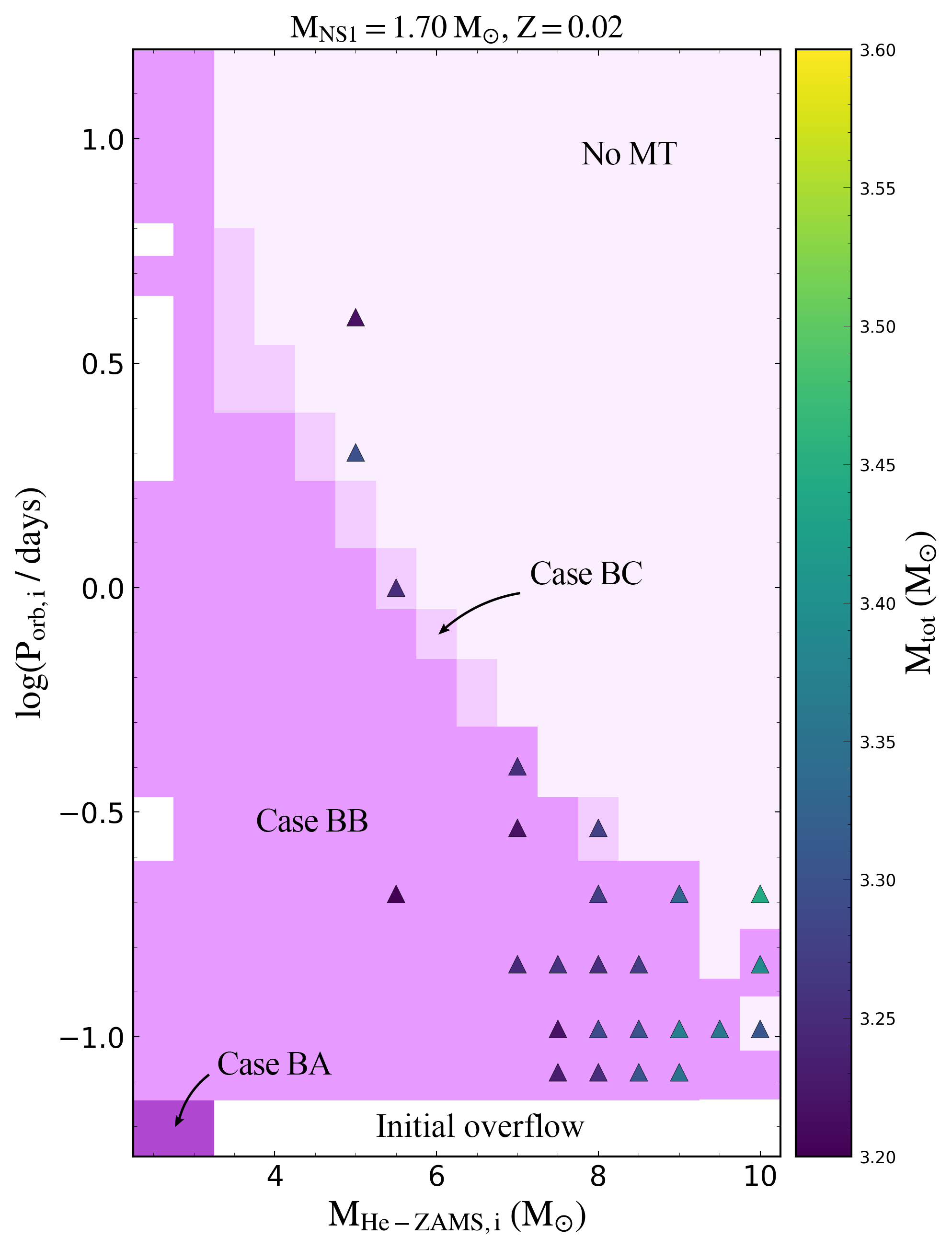}
    \caption{Parameter space of initial orbital period (log scale) as a function of initial He-star ZAMS mass, for a first-born neutron star mass of $M_{NS1} = 1.7$\,\msun{} and metallicity Z = 0.02. The shaded purple background indicates the type of mass transfer (MT) each system undergoes, ranging from Case BA (darkest) through Case BB and Case BC to No MT (lightest), as labelled directly on the panel. White regions mark the initial overflow of the He stars, i.e., the He stars that already overflows its Roche lobe at the beginning of the evolution, and some initially failed models at $M_{\rm{He--ZAMS,i}} = 2.5$\,\msun{}.
    Triangles mark the potential progenitor He star + NS systems of GW190425, whose predicted double neutron star total mass is consistent with the GW190425 measurement. The color bar on the right shows the GW190425 total mass $3.4^{+0.3}_{-0.1}$\msun{}, with the total mass increasing from purple to yellow.}
    \label{fig:gw190425_progenitor}
\end{figure}

Fig.~\ref{fig:formation_channels_frac} shows the fraction of systems that survive in each channel through the consequent phases until the formation of heavy DNSs.
Under our adopted weighting assumptions, particularly with the double Gaussian (DG) fit representative of the first-born neutron star (NS1) masses in observed DNS systems, we find that heavy GW190425-like DNS systems remain rare at solar metallicity. In particular, the standard formation channel produces GW190425-like systems with a total fraction of $\sim$0.5 per cent. While this demonstrates that the standard isolated binary evolution channel is capable of forming such systems, the predicted fraction is still relatively smaller than the detected fraction. 
At solar metallicity, He star models form GW190425-like systems in a narrow mass range of $\sim$4.5\,\msun{}--9.8\,\msun{}. For example, Fig.~\ref{fig:gw190425_progenitor} shows a $P_{\rm{orb}}$--$M_{\rm{He-ZAMS,i}}$ parameter space for a massive NS1 of mass 1.7\,\msun{} at solar metallicity, highlighting the progenitors of systems with total masses consistent with GW190425 that will merge within a Hubble time. 
Most of the GW190425-like systems are formed through Case BB RLOF MT. Our parameter space covering the GW190425 progenitors is broader than the low-spin prior model with $M_{\rm{NS1}} = 1.72$\,\msun{} reported in \citet{Qin:2024A&A}.

The unusually heavy mass of GW190425 and the lack of any similar observation in our Milky Way disk motivated an explanation of the formation of heavy DNSs with alternate formation channels. We find that the formation of heavy GW190425-like systems through intermediate-mass He stars with massive first-born neutron star companions \citep{Qin:2024A&A}, as well as through massive He stars with a broad range of first-born neutron stars \citep{VignaGomez:2021ApJL}, \textit{is rare}. These channels contribute less than 0.0002 per cent and 0.00006 per cent, respectively. According to the adopted IMF-motivated distribution for helium stars, massive helium stars form more rarely than low-mass helium stars. Similarly, massive neutron stars are rare in nature if we believe the Galactic recycled NS population to be the representative of the NS distribution at solar metallicity. Due to this intrinsic nature in our modelling, these channels contribute insignificantly to the formation of heavy DNSs at solar metallicity.

\subsubsection{At low metallicity}
\label{subsubsec:lowZ190425}

\begin{figure}
    \centering
    \includegraphics[width=\columnwidth]{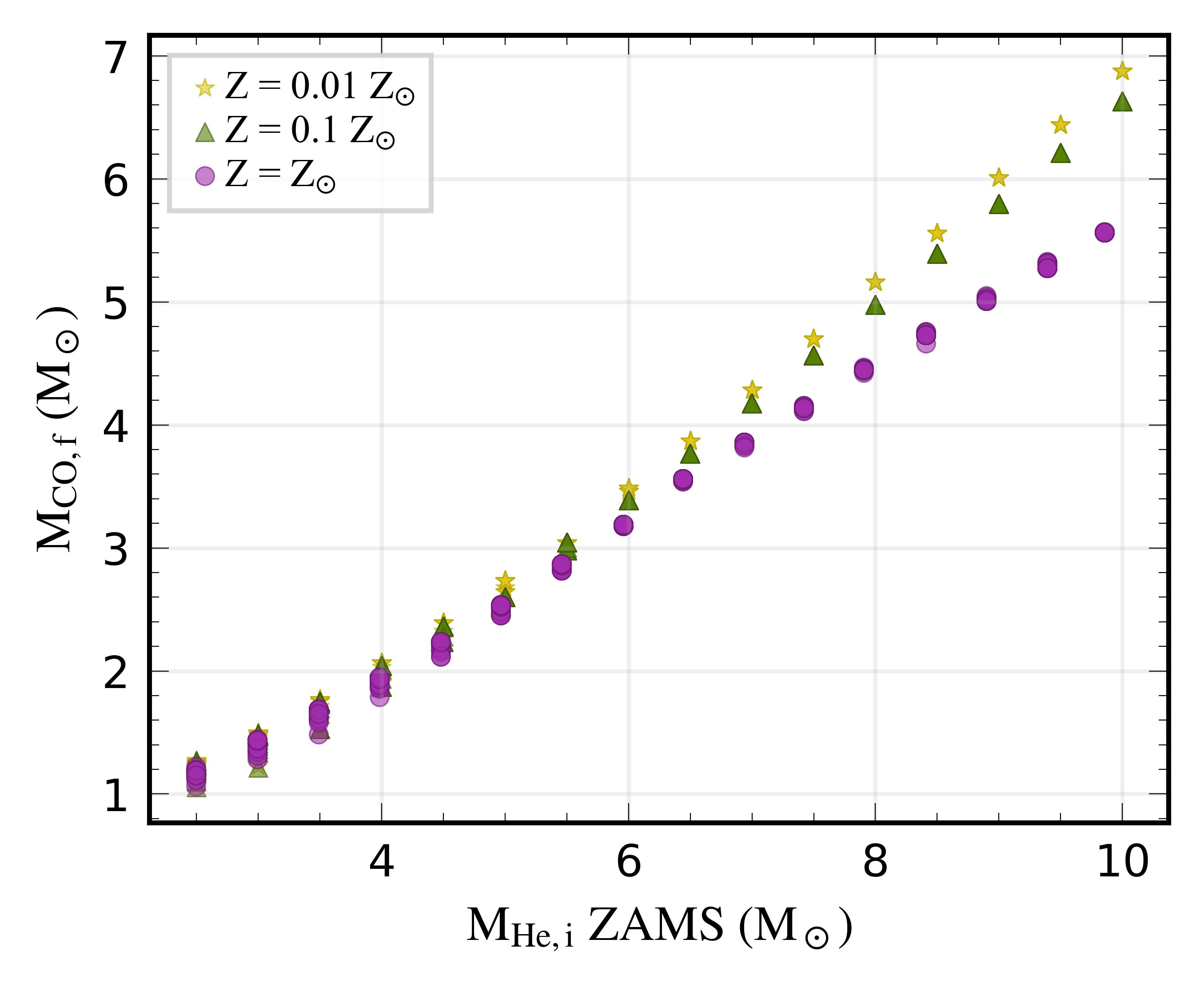}
    \caption{Final (pre-SN) carbon-oxygen (CO) core mass as a function of initial HeZAMS mass for helium stars in He in a binary with an NS1 of mass 1.3\msun{}, with varying orbital periods. Different marker shapes denote different metallicities;
    $Z = \mathrm{Z_{\odot}}$ (purple circle),  $Z = 0.1\,\mathrm{Z_{\odot}}$ (green triangles), and $Z = 0.01\,\mathrm{Z_{\odot}}$ (gold stars).}
    \label{fig:co_core}
\end{figure}

\begin{figure}
\centering
\includegraphics[width=\columnwidth]{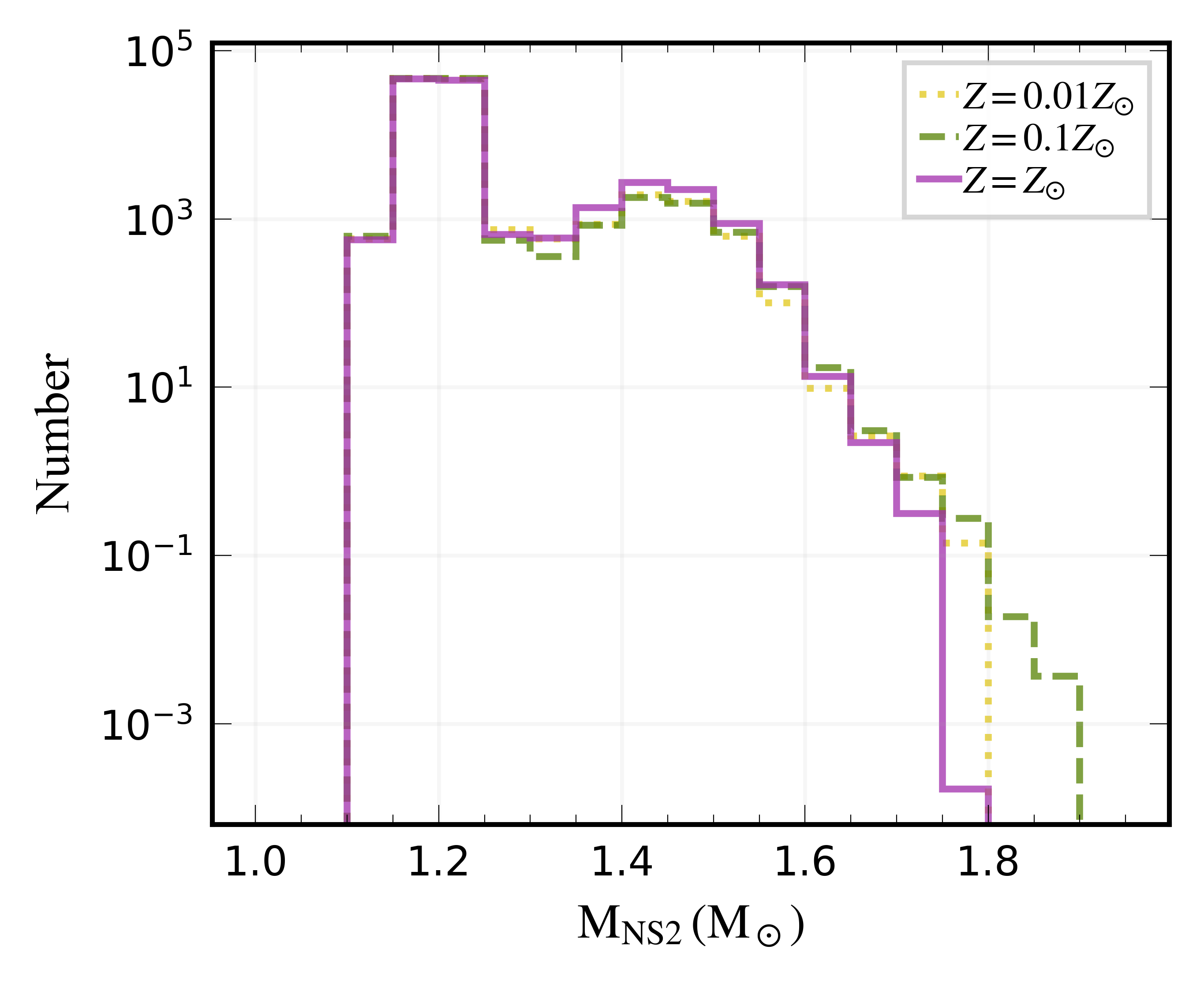}
\caption{Distribution of the masses of second-born NSs (NS2) from the simulation at three different metallicities represented with solid purple, dashed green, and dotted yellow lines corresponding to Z = $\mathrm{Z_{\odot}}$, 0.1\,$\mathrm{Z_{\odot}}$, and 0.01\,$\mathrm{Z_{\odot}}$, respectively.}
\label{fig:all_z_ns2}
\end{figure}

Radial expansion and wind-driven mass loss in massive stars are strongly influenced by metallicity, which in turn affects the pre-SN metal core mass. Fig.~\ref{fig:co_core} shows the pre-SN CO core masses of our He-star models in binaries with NS1 companions at metallicities of $\mathrm{Z_{\odot}}$, 0.1\,$\mathrm{Z_{\odot}}$, and 0.01\,$\mathrm{Z_{\odot}}$. At lower metallicities, He stars experience weaker stellar winds and therefore lose less mass prior to the supernova explosion. As a result, they retain slightly more massive pre-SN metal cores than their solar-metallicity counterparts. Since the remnant neutron star mass is primarily determined by the pre-SN CO core mass in our remnant prescription, we expect lower-metallicity models to produce more massive second-born neutron stars (NS2s). Fig.~\ref{fig:all_z_ns2} shows the resulting NS2 mass distributions for all three metallicities. At 0.1\,$\mathrm{Z_{\odot}}$, NS2 masses range approximately from 1.1--1.85\,\msun{}, while at 0.01\,$\mathrm{Z_{\odot}}$ they range from 1.1--1.8\,\msun{}, with slightly more NS2s compared to those formed at solar-metallicity, reflecting the enhanced mass loss experienced by their progenitors.

\begin{figure}
    \centering
    \includegraphics[width=\columnwidth]{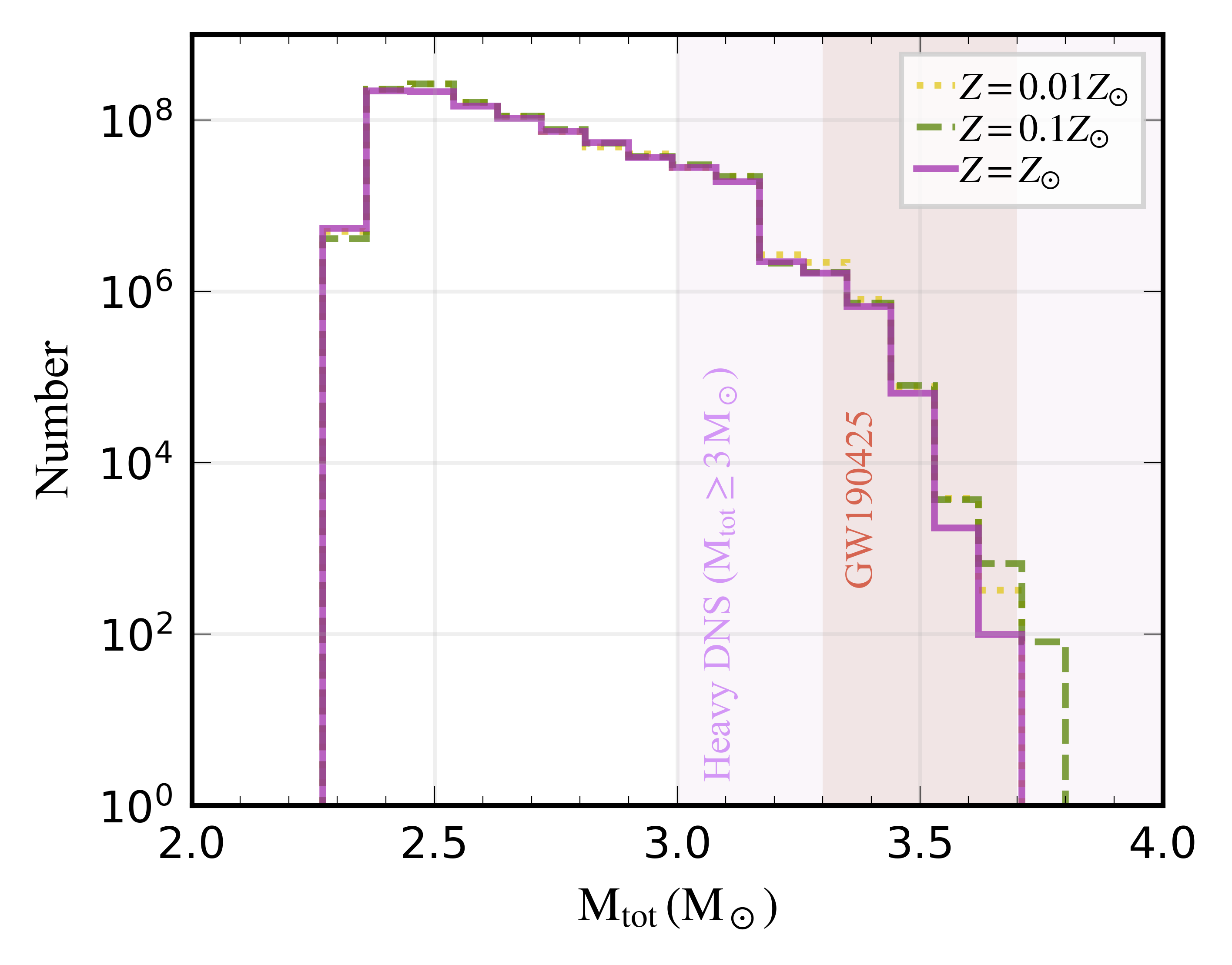}
    \caption{Distribution of total masses ($M_\mathrm{tot}$) of all DNSs at Z = $\mathrm{Z_{\odot}}$, 0.1\,$\mathrm{Z_{\odot}}$ and 0.01\,$\mathrm{Z_{\odot}}$ 
that formed in our simulations. The three different metallicities are represented by solid purple, dashed green, and dotted yellow lines corresponding to Z = $\mathrm{Z_{\odot}}$, 0.1\,$\mathrm{Z_{\odot}}$, and 0.01\,$\mathrm{Z_{\odot}}$, respectively.
The light shaded purple region indicates DNSs with $M_\mathrm{tot}> 3$\,\msun{}, and the shaded red region shows the inferred $M_\mathrm{tot}$ for GW190425 \citep[][]{Abbott:2020ApJL}.}
    \label{fig:dns_hist}
\end{figure}

We compare the fraction of heavy DNSs at solar and low metallicity, focusing on the overall DNS population. Fig.~\ref{fig:dns_hist} shows the total mass distribution of DNSs formed in our models at $\mathrm{Z_{\odot}}$, 0.1$\mathrm{Z_{\odot}}$ and 0.01$\mathrm{Z_{\odot}}$.
To represent the NS1 mass distribution across metallicities formed at different times in the Universe, we adopt the TOP model weights and apply them across all metallicities. 
We find that 4.8 per cent (3.6 per cent) of all systems that become DNSs (of all He star + NS parameter space) are formed heavy at solar metallicity ($\mathrm{Z_{\odot}}$). In comparison, 5 per cent (3.7 per cent) of DNS systems are formed heavy at 0.1$\mathrm{Z_{\odot}}$ and 4.9 per cent (3.5 per cent) at 0.01$\mathrm{Z_{\odot}}$, contributing significantly to the massive DNS population. The figure indicates that the resulting difference between the fraction of total mass of heavy DNSs at low and solar metallicity is small, with only $\sim$0.2 per cent more massive DNSs formed at 0.1$\mathrm{Z_{\odot}}$ and $\sim$0.1 per cent at 0.01$\mathrm{Z_{\odot}}$, compared to the fraction of heavy DNSs at solar metallicity. 
In our simulation, 92 per cent of the DNS systems formed merge within a Hubble time at solar metallicity, with 94 per cent slightly more merging at both 0.1\,$\mathrm{Z_{\odot}}$ and 0.01\,$\mathrm{Z_{\odot}}$. Of the DNSs that merge within the  Hubble time, 4.8 per cent contribute as heavy DNSs at solar metallicity, 5 per cent at 0.1$\mathrm{Z_{\odot}}$ and 4.9 per cent at 0.01$\mathrm{Z_{\odot}}$, indicating that the metallicity has little to no effect on the merger population of heavy DNSs either. 

Our models indicate that the total DNS mass distribution is weakly dependent on metallicity, consistent with previous work \citep[e.g.,][]{vanSon:2025ApJ}, but in contrast to \citet{Giacobbo:2018MNRAS}, where they find a significant fraction of heavy DNSs at low metallicity (Z < 0.008), and \citet{Gallegos-Garcia:2023ApJ}, who see dependence on metallicity.

We do not investigate the sub-channel contributions within the low-metallicity models; however, we find that at low metallicity, heavy DNSs formed through mass transfer are predominantly formed through the unstable formation channel. For example, at 0.1$\mathrm{Z_{\odot}}$, of the 5 per cent heavy DNSs reported above, 4.6 per cent were born through the unstable mass-transfer channel, and the remaining 0.3 per cent were born through the stable mass-transfer channel. It is important to note that we identify systems that undergo potentially unstable mass transfer using the criterion that the peak mass-transfer rate from the He star satisfies $\max\,(\log(|\dot{M}_\mathrm{He}$/\msun{}\,$\mathrm{yr}^{-1}|)$$>$ -1. However, we apply the instability criterion only during post-processing, so systems identified as unstable are not evolved through CE phase. Some systems identified as undergoing unstable mass transfer may instead merge during the subsequent CE phase, which is not modelled in this work. The quoted fraction therefore represents the fraction of binary systems that transfer mass and satisfy our instability criterion, rather than the fraction of systems that definitely survive this phase and form DNSs.

\section{Discussion}
\label{sec:discussion}

\subsection{Why does heavy DNS total mass distribution stay unaffected by metallicity?}
\label{subsec:discussion_a}

\begin{figure}
    \centering
    \includegraphics[width=\columnwidth]{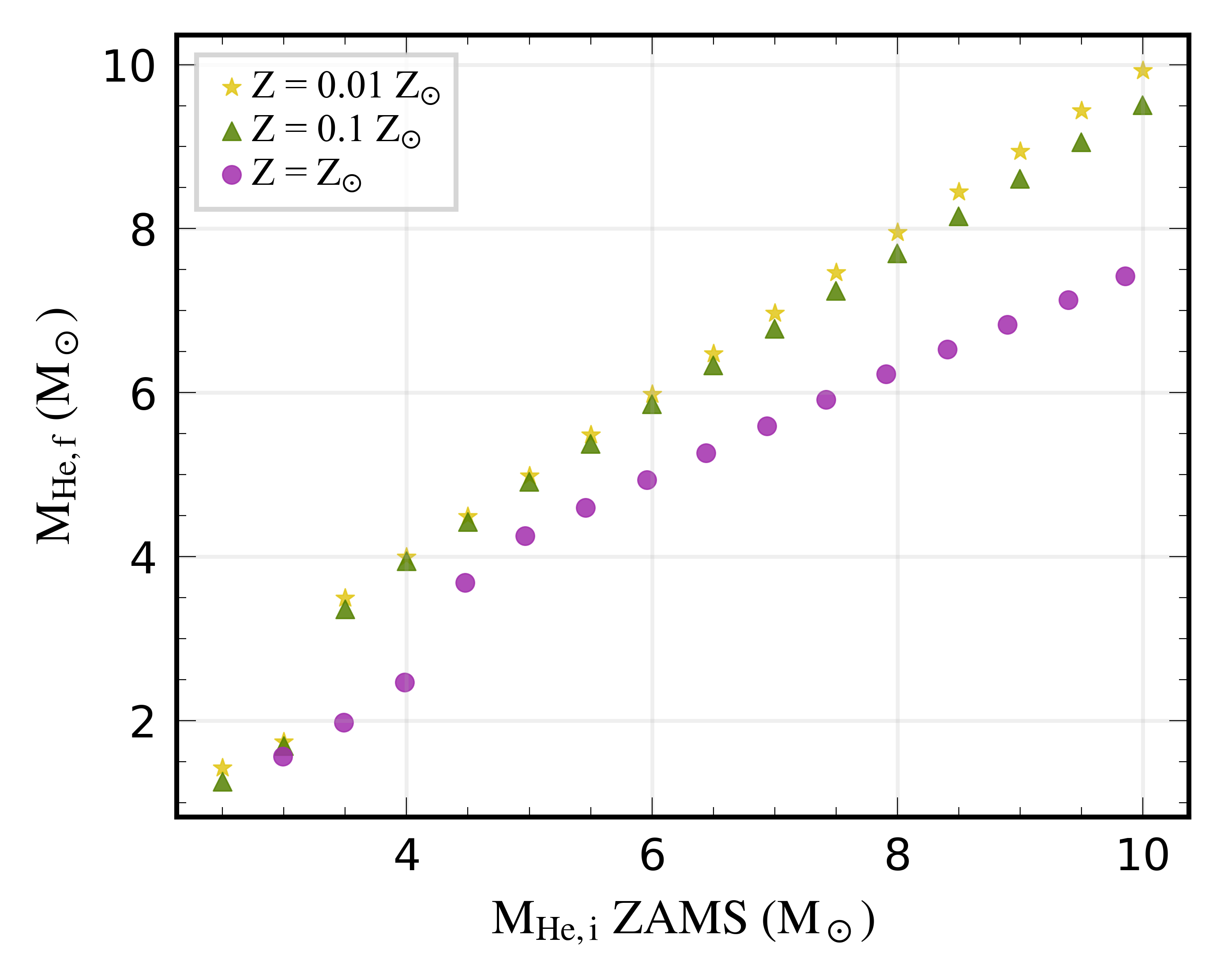}
    \caption{Final He-star mass ($M_\mathrm{He, f}$) as a function of initial HeZAMS mass ($M_\mathrm{He,i}$) for He stars in a binary with NS1 of mass 1.3\msun{}, with varying orbital periods. Different marker shapes denote different metallicity, circle corresponds to Z = $\mathrm{Z_{\odot}}$, triangles to 0.1\,$\mathrm{Z_{\odot}}$, and star to 0.01\,$\mathrm{Z_{\odot}}$ }
    \label{fig:final_he_star_mass}
\end{figure}

Firstly, the He star evolution plays an important role in this result. Prior to Roche-lobe overflow, He stars lose mass through stellar winds, with the effect being stronger at higher metallicity. For example, Fig.~\ref{fig:final_he_star_mass} shows the resulting final helium star masses as a function of initial mass for Z = $\mathrm{Z_{\odot}}$, 0.1\,$\mathrm{Z_{\odot}}$ and 0.01\,$\mathrm{Z_{\odot}}$ (mention the orbital period). After mass transfer, the systems form pre-SN CO cores, which differ by only $\sim$$0.2$--$0.5$\,\msun{} between solar and low-metallicity models (see Fig.~\ref{fig:co_core}).

In the adopted \cite{Mandel:2020MNRAS} prescription, the mapping between pre-supernova CO core mass and the remnant is loosely correlated and approximated by a broken linear fit with stochastic scatter, allowing massive cores to form either neutron stars or black holes depending on the exact core structure. 
Low-metallicity helium stars tend to produce slightly more massive cores, which are more likely to collapse into low-mass black holes within this prescription, whereas solar metallicity models more often remain within the neutron star–forming regime (see~Fig.\ref{fig:co_core}). Hence, the overall effect on the massive DNS population remains small.

Secondly, it is important to emphasise that our models were evolved from the post-CE phase consisting of a He star + NS binary, assuming that the systems have survived the CE phase. We do not model the earlier evolution of the hydrogen-rich (H-rich) OB ZAMS progenitors evolving up until the CE phase. 

The impact of metallicity is different on the evolution of hydrogen stars and helium stars. The evolutionary phases we do not model, such as the hydrogen main-sequence (H-MS), post-main-sequence (post-MS), and the mass transfer episode from the H-rich stars, are likely where the dominant metallicity-driven differences in binary evolution are reflected. 
During the H-rich phase, metallicity strongly modifies stellar radial expansion, the age at which the star would undergo RLOF and transfer mass, thereby affecting how long the envelope stripping lasts. \citet{Klencki_2020} discusses the role of metallicity in the context of massive binary star evolution, which is discussed in their sec. 3.2 and 5 (see also their fig. 3). According to their study, at solar metallicity, RLOF occurs shortly after the MS phase, when the star is evolving through the Hertzsprung gap (HG). The star's envelope expands rapidly, and mass transfer begins relatively early, which allows the stripping of the envelope to last longer.
 
At low metallicity, stars do not expand enough to overflow the RL immediately after the MS phase. By the time RLOF begins, the star is burning He in its core. Stripping of the envelope occurs later during core-He burning, allowing the He core to grow through prolonged hydrogen-shell burning. This shortens the duration of the stripping of the star. The uncertainties associated with the common envelope evolution add to the puzzle of the formation of DNSs \citep[][]{Giacobbo:2018MNRAS, vanSon:2025ApJ}. 

These differences can significantly affect several properties of the He star entering the He star + NS grid, such as the internal structures and radial evolution, compared to He stars at solar metallicity. This will significantly alter the mass and orbital separation parameter space of He star + NS binaries that become heavy DNSs.

After the hydrogen is stripped, and the star is He-rich, the metallicity variation can affect the hydrogen mass fraction retained onto the helium core \citep[][]{gotberg2023stellarpropertiesobservedstars}, which can further affect the mass that the He star loses via wind loss, radius evolution, mass transfer cases boundaries, final CO core mass (see Fig.~\ref{fig:co_core}), remnant mass (see Fig.~\ref{fig:all_z_ns2}), all of which we see in this work. The metallicity effect is important, but it might still be weaker in structurally modifying the He star than in an H-rich star.

In order to investigate the impact of metallicity on DNS formation from the zero-age main sequence, we use the rapid population synthesis code \texttt{COMPAS} \citep[][]{Stevenson_2017, Vigna-Gomez:2018MNRAS, Riley_2022, compas2025} to study the population of DNSs formed at a broad range of metallicities by evolving hydrogen-rich OB-type stars until they form DNS systems. 
The mass-loss rate prescriptions used in COMPAS are described in \citet{Merritt:2026ApJ}. 
Fig.~\ref{fig:COMPAS_sim} shows the total masses for a population of DNSs evolved at different metallicities in COMPAS.
With the default COMPAS assumptions, we find that the total mass distribution and formation efficiency of double neutron stars do not show a significant dependence on metallicity, even with this full-scale binary evolution population.
In the COMPAS population, around $20\%$ of double neutron stars are heavy double neutron stars with \mtot{}\,$\geq 3$\,\msun{}, consistent with previous works using COMPAS \citep[e.g.,][]{Chattopadhyay:2020MNRAS}.


As a result, the weak metallicity dependence we observe may reflect either (i) the fact that the dominant metallicity-driven differences arise earlier in binary evolution, before the He star + NS phase we model, or (ii) a genuine weak dependence of metallicity on heavy DNS formation.

\subsection{Comparison with other studies}
\label{subsec:discussion_b}

In order to explain the formation of heavy DNSs ($M_\mathrm{tot} \ge3$\,\msun{}) such as GW190425, models must simultaneously explain the absence of heavy systems in the observed Galactic field population. There has been growing effort to understand the underlying formation mechanism, the physical processes and properties that drive this discrepancy between the Galactic field and GW population.

As mentioned in Sec.~\ref{sec:methods}, we use a detailed binary evolution code to evolve our binaries from the post-CE phase, which differs methodologically from other binary population synthesis (BPS) codes and can produce obvious differences in the predicted population. Different BPS codes adopt different assumptions regarding stellar wind mass loss rates, CE efficiency, kick mechanism and its dependence on supernova type, and remnant-mass prescriptions, each of which shifts the predicted fraction of heavy DNSs. A one-to-one comparison against every study in the literature is not possible, but we highlight a few studies that are closely related to our work here.

\begin{enumerate}
    
\item Across our simulated models, the total fraction of heavy DNSs ($M_\mathrm{tot} \ge3$\,\msun{}) in the population, representative of the GW merger population, is 0.8 per cent, while the subpopulation of recycled systems is 0.7 per cent, formed with models at solar metallicity. 

\citet{Chattopadhyay:2020MNRAS} studied the Galactic DNS population using the binary population synthesis code COMPAS \citep[][]{Stevenson_2017,Riley_2022} and found that $\sim 30$ per cent of DNS had $M_\mathrm{tot} \ge 3$\,M$_\odot$, in tension with the observed Galactic pulsar population.
A recent binary population synthesis study by \citet{Chu:2025ApJ} using a modified BSE \citep[][]{Hurley:2002MNRAS} code similarly found that 19--22 per cent of merging DNSs at redshift $z = 0$ are heavy, with $M_\mathrm{tot} > 3$\,M$_\odot$.
\citet{Chu:2025ApJ} argue that the heavy DNSs in their model are less represented in the radio pulsar population because the pulsar is less recycled (has a higher magnetic field and a longer spin period) than those in low-mass DNS.
There are a number of possible reasons that these studies find a greater fraction of heavy DNSs than we do.
Firstly, these studies use binary population synthesis codes, where the initial binary population is sampled from an assumed initial mass function, mass-ratio distribution, and orbital-separation distribution and evolved from zero-age main sequence, implying potentially non-trivial distributions for NS1, the He-star mass, and the binary orbital period, as well as allowing for formation channels that our work does not model, such as the double core common envelope channel \citep[][]{Dewi:2006MNRAS,Vigna-Gomez:2018MNRAS}. In contrast, our work focuses on modelling the final evolutionary phases in detail, but with simple parametric distributions assumed for the initial distributions for NS1, the He star mass and the initial orbital period. 
Secondly, we do not account for radio or GW selection effects \citep[e.g.,][]{Os_owski_2011, Chattopadhyay:2020MNRAS,Sgalletta:2023MNRAS} which can lead to biases in which DNSs are observable. 
However, previous work has shown that these do not change the DNS mass distribution dramatically. 

\item To check if the disagreement between our study and \citet{Giacobbo:2018MNRAS} on the formation of massive DNSs at low metallicity is due to the modelling differences, we try two modelling variations:

 a. \textit{Wind mass loss prescription:} Initially, we modelled the mass loss of He stars using the same recipe for the solar metallicity models, i.e., the default MESA \citet{2000A&A...360..227N} prescription (hereafter, NL2000). We modelled our low-metallicity models by adopting a metallicity-dependent wind mass loss treatment by \citep[][hereafter VD2005. See Ref.~\ref{sec:methods} for details]{2005A&A...442..587V}. As seen in Fig.~\ref{fig:NL_VD_Mhe_i_vs_f_Porb0.145d} and Fig.~\ref{fig:NL_VD_MCO_i_vs_f_Porb0.145d}, the final He star mass and the final CO core of the He star, respectively, only differ approximately by $\leq 0.5$\,\msun{}. Even though the properties of He stars in a binary evolution with NSs showed the effect of winds due to metallicity, it did not make a significant difference in the final DNS total mass distribution; that is, our models did not form a larger fraction of heavy DNSs as observed in \citet{Giacobbo:2018MNRAS}.

The wind-mass loss rates from He stars and Wolf--Rayet (WR)-like stars remain weakly constrained. There are various treatments used in the literature to model wind mass loss from He stars, especially to incorporate metallicity dependence. 
For example, for their stripped He stars, \citet{G_tberg_2017} empirically extrapolate the wind mass loss prescription of \citet{2000A&A...360..227N}, where the prescription has $\beta = 0.56$. \citet{Vink_2017} focused on making predictions on stripped helium stars (for a $\sim$\,4\,\msun{} He star) with a Monte Carlo radiative transfer model and came up with an updated mass loss rate, where the predicted exponential factor of metallicity dependence $\beta$ is 0.61, which results in mass loss rate of an order of magnitude less than that predicted in \citet{G_tberg_2017}. The predicted value is a little smaller than what we use in this work based on the prediction by \citet{2005A&A...442..587V}. On this, \citet{Vink_2017} mentions that it is the dependence of the assumed wind velocity at infinity $v_{\infty}$ on Z. This is the closest theoretical prediction we have for low/intermediate mass stripped helium stars. Other recent predictions for massive helium stars (WR stars) are given by \citet{Sander_2020}. From the recently discovered intermediate stripped helium stars in the range 2--8\,\msun{} \citep{Drout:2023Sci}, \citet{gotberg2023stellarpropertiesobservedstars} provided a detailed constraint on the wind mass-loss rate from these stripped helium stars. Fig. 12 of their paper shows a comparison of mass-loss rates from their observation to various known He-star wind mass-loss recipes mentioned above, showing that their estimates of mass-loss rates from observation are lower than all of those prescriptions. Additionally, recent studies present a detailed study and comparison to Wolf--Rayet wind mass-loss rates using different prescriptions
\citep[for e.g., see][]{Higgins_2021, 2025A&A...697A.114P}.
This sums up the uncertainty in constraining mass-loss rates from helium stars in the range that we have modelled in our work. Further, it clearly points to the fact that having observations like those reported in \citet{Drout:2023Sci} of intermediate-mass helium stars that become neutron stars at the end stage of their evolution will be helpful in constraining wind mass-loss rates, which iin turn impacts the predictions of compact binary evolution. It is out of scope to compare each of these recipes to demonstrate how they affect the final fraction of heavy DNSs, but the comparative study mentioned in this work highlights that it would not significantly alter the He star in the He star + NS evolutionary phase.

b. \textit{Remnant mass calculation and kick velocity:} As discussed in Sec.~\ref{subsec:discussion_a}, we use the \citet{Mandel:2020MNRAS} prescription for calculating the post-SN remnant mass and natal kick velocity magnitude of the NS2. The kick velocity in this study is a physically motivated approach.
Whereas \citet{Giacobbo:2018MNRAS} uses a rapid core-collapse SN model, a SN mechanism introduced by  \citet{2012ApJ...749...91F}, and draws the natal kick velocity of newborn NSs (NS2) from the Hobbs-Maxwellian distribution \citep{Hobbs_2005}. This is a purely statistical approach. 
Both prescriptions calculate remnant masses from the pre-SN CO core masses.
Naturally, to check if the underlying differences in the final fraction of massive DNSs formed through our models and their study were due to the choice of remnant prescription and the way natal kick velocity is modelled, we simulate our models following the same approach used in \citet{Giacobbo:2018MNRAS}. Fig.~\ref{fig:fryer_dns_hist} shows that, even after incorporating these changes, the DNS total mass distribution is not significantly affected by metallicity. We form fewer massive DNSs at low metallicity than models at solar metallicity when using the rapid core-collapse SN remnant prescription. This is because the massive He-star models with initial ZAMS masses of $\geq 9.5$\,\msun{} produce CO cores with masses of $\geq$\,6.1,\msun{}, which, according to the rapid core-collapse SN remnant prescription, undergo complete fallback and form black holes. In contrast, at solar metallicity, the models produce CO core masses that lie well within the NS formation range. This difference significantly affects the resulting DNS total mass distribution.
As we do not observe this behaviour in \citet{Giacobbo:2018MNRAS}, this could indicate that the He-star--NS parameter space that forms DNSs at different metallicities in their work may vary. However, we do not recover the expected trend with metallicity. Therefore, we rule out the choice of remnant prescription as the primary cause of the difference in the low-metallicity DNS total mass distribution. Additionally, the difference could stem from the fact that we consider an independent parameter grid of He-ZAMS star masses and initial orbital periods, assuming that all systems survive the post-CE phase. As discussed in the previous Sec.~\ref{subsec:discussion_a}, the parameter space of He star -- NS binaries with a mass range of He-rich stars and orbital period, that enter the parameter space to form heavy DNSs might vary depending on the evolution of hydrogen ZAMS binaries.

However, in fig. 2 of \citet{Giacobbo:2018MNRAS}, the metallicity dependence of the heavy DNS population vanishes for the \texttt{CC15$\alpha$5} model, which assumes a low natal kick for core-collapse supernovae. This indicates that the effect might be partially a consequence of the kick velocity prescription, making natal kick velocity as one of the key factors shaping the total mass distribution of DNSs. Additionally, the differences might also hint at differences in other physical assumptions such as the overshooting parameter, which we do not model.

\item A recent study by \citet{Chattaraj:2026ApJ} found that DNSs in their models are predominantly formed via the CE channel, but they exclusively showed that the CE channel actually splits into two distinct sub-channels, namely the He core and C/O core channels. The He core and C/O core channels are distinguished by the evolutionary stage of the donor at the onset of CEE, depending on whether the donor has a helium core or has already developed a C/O core. The study claim that the He core and C/O core channels map onto the two observed Galactic DNS sub-populations, split at roughly $\sim$ 1\,d: the He core channel produces the short-period systems ($\sim$ 0.08--0.6 d) that merge within the Hubble time, while the C/O core channel produces the longer-period systems (sub-population ii, $\sim$1--19 d) that do not merge within Hubble time. We test if the model parameter space we simulate reproduces this bifurcation. For the He core channel, we select all of our He star + NS star models as before. For the C/O core channel, we select models that avoid RLOF on the He main sequence (HeMS), such that they only interact after they have developed a C/O core. 
In our Fig.~\ref{fig:he_co_core_bifurcation},
we see a bifurcation of these two channels at $\sim 1$ d, indicating that our model agrees with this claim. However, with our modified prescription, the wide systems are disrupted, leaving us with fewer systems contributing to the C/O core channel.

In the same study, \citet{Chattaraj:2026ApJ} introduced a new physically motivated kick mechanism for the natal kick imparted onto the newborn NS, namely the `Asymmetric Ejecta' kick model. This approach uses the gravitational tug-boat mechanism for asymmetric supernova explosion introduced by \citet{Janka_2017}. In our models, at solar metallicity, we can explain the formation of most observed DNSs with small kick velocities produced by ECSN and mainly USSN, CCSN with a He star progenitor of $\sim$3--3.5\,\msun{}, except for the systems at high eccentricities, which need a high kick velocity from a He star progenitor of mass $\sim$4\,\msun{} to explain their formation. Using this new physics-based kick prescription would be useful to check if our models can reproduce the observed systems with a lower kick velocity of $\leq$40\,$\rm{kms^{-1}}$.

\item We stop our models at core carbon depletion. At this point, our low-mass He star models with masses 2.5--3\,\msun{} undergo only the second mass transfer phase. However, \citet{WuFuller:2022ApJL} recently showed the possibility of late-time mass transfer from 2.5--3\,\msun\ helium stars during oxygen/neon burning in the final $\sim$10\,yr before core collapse \citep[see also][]{Jiang:2021ApJL}. 
This late-time mass transfer could significantly alter the binary properties prior to core collapse.

\end{enumerate}

\section{Conclusions}
\label{sec:conclusions}

The unusually heavy mass of the gravitational-wave source GW190425, a DNS detected during the O3 run, has raised a question about its formation. This uncertainty has motivated the research community to explore alternate formation pathways to explain its formation. However, these alternate pathways are explained using several physical assumptions independently, resulting in an unclear picture of which of these formation channels is more likely to form a heavy GW190425-like DNS (with $M_\mathrm{tot} \ge 3$\,\msun{}) under a self-consistent framework.
In this work, we aim to test the formation channels mentioned in Sec.~\ref{sec:intro} by building a self-consistent framework.

One of the commonly known caveats with binary population synthesis codes is that they fail to reproduce and explain the observed population of Galactic DNSs in the field while explaining the rate of GW190425. 
In Paper I, we aimed to explain the observed population of DNSs in our Galactic field with our simulated DNSs. We went about this by simulating the He star + NS binaries using a detailed stellar evolution code, MESA. We simulated a range of He-star masses between 2.5--10\,\msun{}, and a fixed NS mass of 1.4\,\msun{} at solar metallicity ($Z = Z_\odot = 0.02$) and showed that we could somewhat explain the observed population. 
However, the parameter space explored was limited and was not sufficient to fully compare our simulated population to the observed population. Additionally, this restricted parameter space did not allow us to test all the uncertainties associated with the formation of heavy DNS. Therefore, in this work, we extend the immediate progenitor parameter space for DNS formation with He star mass ranging between 2.5--10\,\msun{}, NS1 masses ranging between 1.1--1.9\,\msun{}, both in a binary with orbital periods ranging from 0.06 - 100 d. We model these He star $+$ NS binaries with varying orbital periods at three different metallicities - $Z = Z_\odot = 0.02$ (solar metallicity), 0.1\,$\mathrm{Z_{\odot}}$, and 0.01\,$\mathrm{Z_{\odot}}$.

We adopt the same assumptions and framework used in the first part of this series, Paper I, to model the post-SN properties of our binaries, like the remnant masses of the newborn NS (NS2), natal kick velocity magnitude, and the merging time-scale of the binaries.  
We find that in considering a broader parameter space, the resulting DNS population remains broadly consistent with observations (see~Fig.\ref{fig:interpolated_porb_e} and Fig.~\ref{fig:dns_hist}). We fail to reproduce the observed gap between the low and high eccentricity at short orbital periods \citep[][]{Andrews:2019ApJ, grichener2026bimodalityeccentricitydistributiongalactic}. 
However, we can explain the three systems with intermediate helium stars. 
In the DNS total mass distribution, we marginally overproduce systems with total mass 2.55\,\msun{}.

With these simulated DNSs on a broader parameter space, we calculated the fraction of heavy DNS ($\geq$ 3\,\msun{}) formed in each of the four formation pathways as mentioned in Sec.~\ref{sec:results}. 
We find that :

\begin{enumerate}
    
    \item  Even with a broader parameter space, we do not form enough heavy GW190425-like DNSs at solar metallicity to be able to explain the observed high rate of GW190425.

    \item  At solar metallicity, the standard formation channel dominates with a total fraction of 0.5 per cent of heavy DNSs as compared to an insignificant fraction of DNSs, $6\,\times 10^ {-5}$ formed through the massive He star channel \citep[][]{VignaGomez:2021ApJL}, and $2\,\times 10^{-4}$ formed through the massive NS1 channel \citep[][]{Qin:2024A&A}. The insignificant fraction of DNSs formed through the later channels is simply because \textit{massive He stars and massive first-born NS (NS1) are rare}. Most importantly, we rule out the formation of heavy DNSs through the unstable formation channel even with massive NS1 at solar metallicity.
    
    \item  Within our models, we do form massive NS2 remnants from He star models at low metallicity; however, such systems remain rare (see Fig.~\ref{fig:all_z_ns2}). We find that the maximum NS2 mass changes very little with metallicity. Overall, we find a weak dependence of heavy DNS total mass on metallicity. The low-metallicity model with $Z = 0.1 $\,Z$_{\odot}$ produces a 0.5 per cent higher fraction of GW190425-like systems than the solar metallicity model, whereas the fraction at $Z = 0.01 $\,Z$_{\odot}$ remains mostly unchanged. It is important to note that our models are not weighted by the cosmic star formation history or the metallicity distribution of star-forming environments. The weak metallicity dependence found in this work could indicate that heavy DNS formation is weakly dependent on metallicity, consistent with previous studies. Alternatively, it may reflect the scope of our modelling, which begins from the He star + NS phase. In that case, much of the metallicity sensitivity that might be experienced during the wind mass-loss rates and earlier mass transfer episodes may have been largely established during the earlier evolution of the progenitor binary when the star is evolving through the hydrogen post-MS phase, thereby impacting which systems ultimately enter the He star + NS parameter space. 
    
    \item The heavy DNS systems at low metallicity are formed through unstable mass transfer, in contrast to the heavy DNS systems at solar metallicity. 
    
    \item This highlights that maybe we do not need an alternate formation channel other than the \textit{standard formation channel} \citep{Tauris:2017ApJ} to explain the formation of a GW190425-like heavy system. 
    In our models, heavy DNSs are simply the high-mass tail of the standard DNS population. 

    \item If near-future GW observations continue to find a large fraction of heavy DNSs, whilst they remain rare in the Galactic population, it may be that a significant fraction of DNSs are born heavy \citep[][]{Chattopadhyay:2020MNRAS,Mandel:2021MNRAS,Chu:2025ApJ} and radio selection effects simply make it challenging to observe heavy DNSs in the Milky Way. 

\end{enumerate}
Recently, observations at different wavelengths have been probing different evolutionary stages leading to the formation of DNS, such as supernova explosion \citep[][]{2018Sci...362..201D, 2023A&A...675A.201A, 2024ApJ...969L..11D, 2024ApJ...965...51B}, high-mass x-ray binary (HMXB) phases \citep[][]{Vinciguerra:2020MNRAS, 2025arXiv250508857V}, and pulsar-He star binary formed post-CE evolution \cite{Yang:2025Science}. 
More such observations in the future can help constrain or rule out formation channels that overproduce or underproduce merging systems.

\section*{Acknowledgements}

AN thanks Alejandro Vigna-Gómez for an insightful conversation. This research was supported by the Australian Research Council (ARC) Centre of Excellence for Gravitational Wave Discovery (OzGrav), through project numbers CE170100004 and CE230100016. 
SS and AN are supported by an ARC Discovery Early Career Research Award (DE220100241; PI Stevenson). 
This work was performed on the OzSTAR national facility at Swinburne University of Technology. 
The OzSTAR program receives funding in part from the Astronomy National Collaborative Research Infrastructure Strategy (NCRIS) allocation provided by the Australian Government, and from the Victorian Higher Education State Investment Fund (VHESIF) provided by the Victorian Government.

\section*{Data availability}

MESA version: \texttt{23.05.01} inlists for He star + NS binary evolution and 
Postprocessing scripts to simulate DNS formation can be found here: 
\href{https://github.com/nair-ashwathi/BinaryNeutronStar_Formation}{GitHub repository}

Simulations in this paper made use of the COMPAS rapid binary population synthesis code (\texttt{v03.29.05}), which is freely available at http://github.com/TeamCOMPAS/COMPAS.




\bibliographystyle{mnras}
\bibliography{bib} 



\appendix

\section{Modelling Helium Zero Age Main Sequence}
\label{sec:appendix_a}
The default stopping condition for the zero-age main sequence (ZAMS) is defined as the point at which the ratio of energy generation from nuclear reactions in the interior of the star to its total luminosity, that is, all energy leaving the star per second, exceeds $>0.99$ ($L_\mathrm{nuc}/{L} > 0.99$). This fraction assumes that the star is considered to reach the ZAMS stage once 99 per cent of its luminosity is nuclear-powered. This condition is well suited to stars at $Z = 0.02$. However, for stars at lower metallicity,
the reduced opacity leads to faster contraction, making the criteria less reliable. We therefore adopt a helium-burning luminosity criterion to define the He-ZAMS, following the method from \citet{Aguilera_Dena_2022}. Rather than relying on the $L_\mathrm{nuc}/{L}$ ratio, this criterion identifies the He-ZAMS as the point at which core He burning comes to dominate the star's energy output. The model is considered to have reached the He-ZAMS once the total energy generation rate from the helium fusion in the star's core accounts for at least 90 percent of the star's luminosity.

\section{NS1 interpolation}
\label{sec:ns1_interpolation}

We use the \texttt{scipy.interpolate.interp1d} class to perform one-dimensional linear piecewise interpolation to estimate CDF for NS1 values that we did not explicitly model. 

\begin{figure}
    \centering
    \includegraphics[width = \columnwidth]{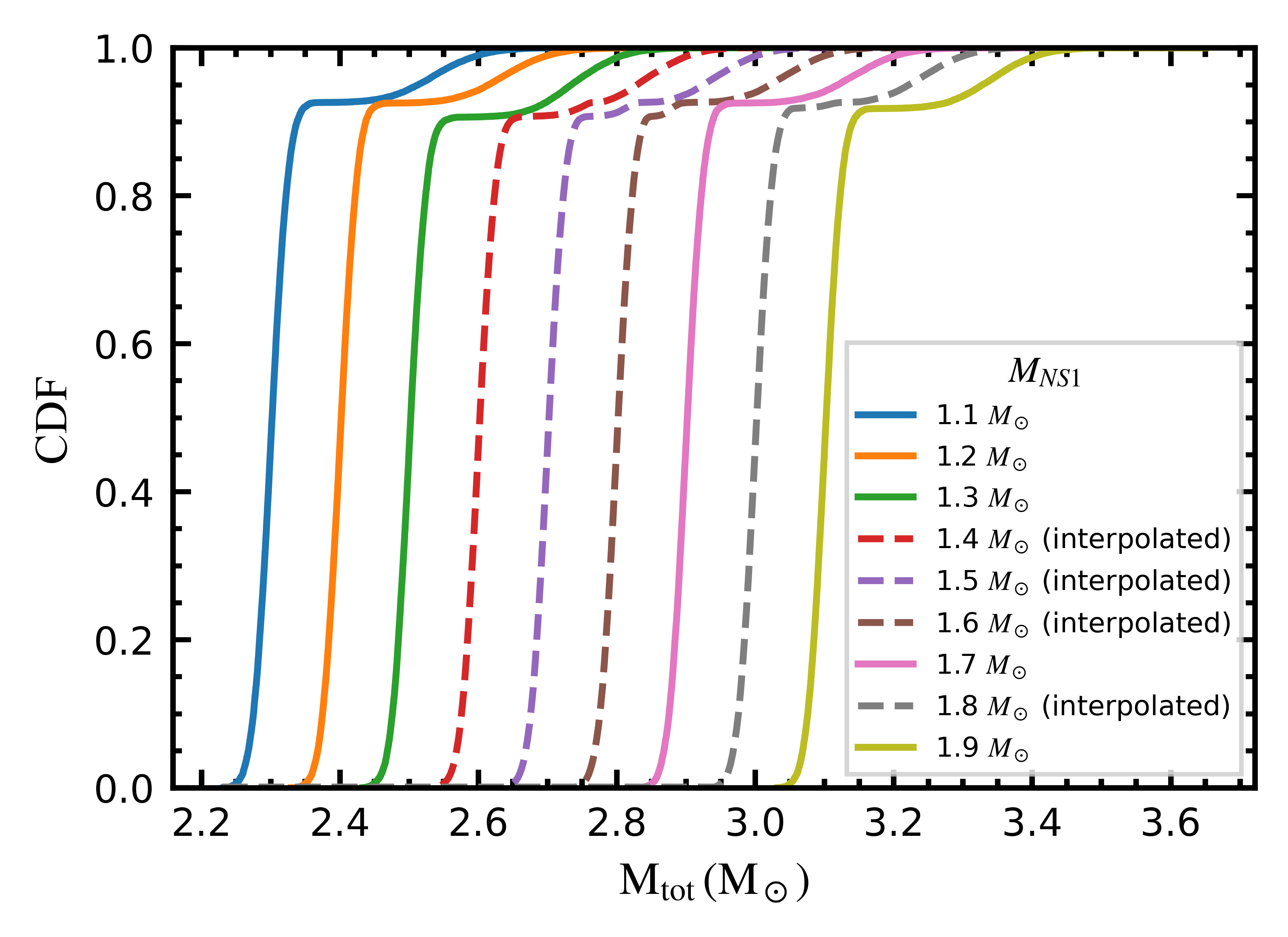}
    \caption{Cumulative distribution functions (CDFs) of the total double neutron star (DNS) system mass (\mtot{} = \mnsone{} + \mnstwo{}), shown for fixed primary neutron star mass \mnsone{} ranging from 1.1--1.9 in steps of 0.1\,\msun{}. Solid lines correspond to the simulated \mnsone{}, whereas the dashed lines correspond to the CDFs of the interpolated \mnsone{} between the simulated ones. With the increase in \mnsone{}, each CDF is systematically shifted toward higher \mtot{}, reflecting the direct dependence of total mass on the first-born NS.}
    \label{fig:interpolated_cdfs}
\end{figure}

\section{Comparison to other studies}

The plots presented in this section support the comparisons with previous studies discussed in the main Sec.~\ref{sec:discussion}; please refer to the same for more details.

\begin{figure}
    \centering
    \includegraphics[width=\columnwidth]{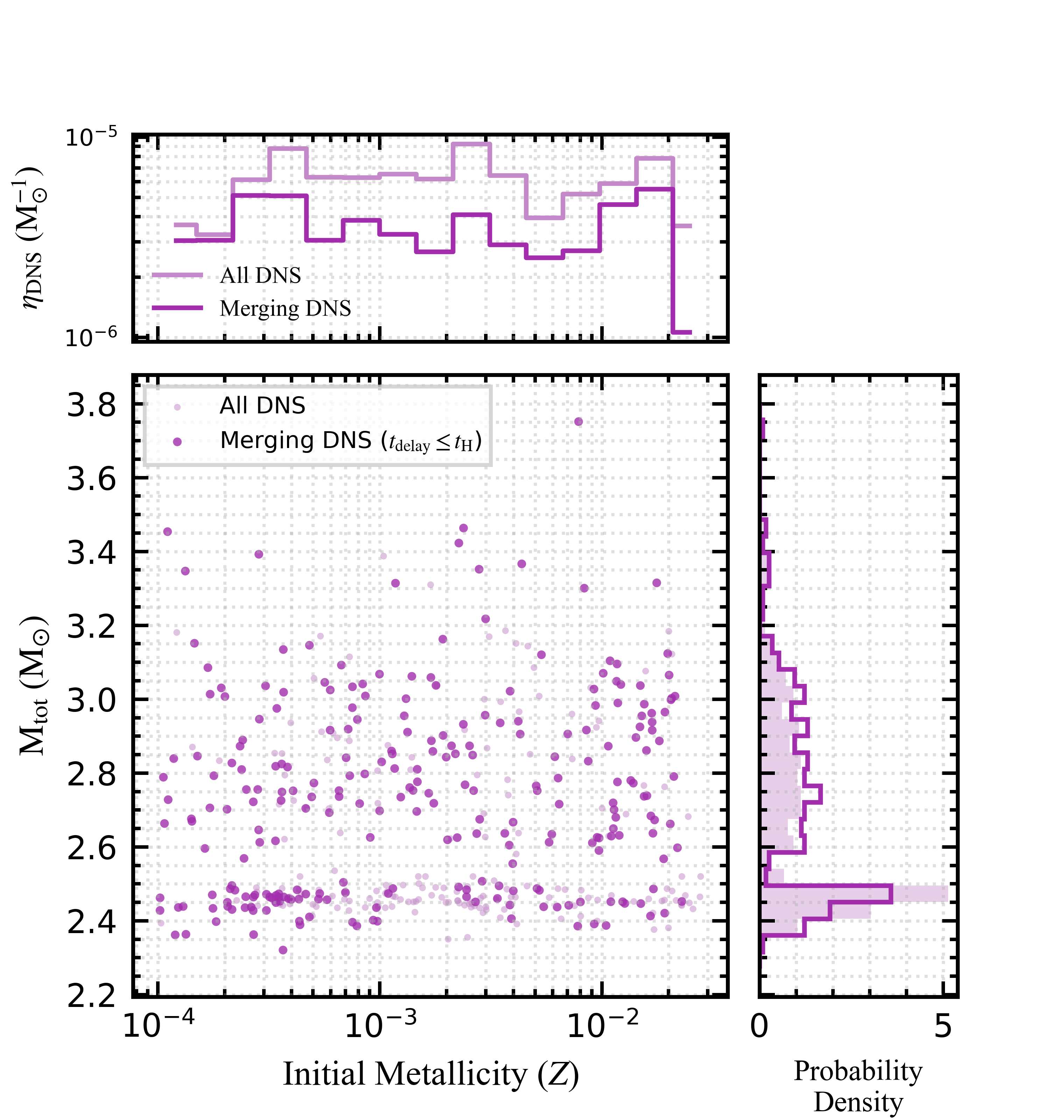}
    \caption{Double neutron star (DNS) properties as a function of initial metallicity ($Z$) from COMPAS binary population synthesis models, based on a simulated population of $2\times10^{6}$ binaries. \textit{Main panel:} total DNS mass, $M_{\rm tot} = M_1 + M_2$, as a function of initial metallicity. Light points correspond to all DNS formed in the simulation; dark points correspond to the subset of systems that merge in the Hubble time.
    \textit{Top panel:} DNS formation efficiency, $\eta_{\rm DNS}$, defined
    as the number of DNS formed per unit star-forming mass,
    shown for all DNS (light) and merging DNS (dark) across the same
    metallicity bins as the main panel.
    \textit{Right panel:} marginal probability density of $M_{\rm tot}$ for
    all DNS (shaded) and merging DNS (outlined step)}
    \label{fig:COMPAS_sim}
\end{figure}

\begin{figure}
    \centering
    \includegraphics[]{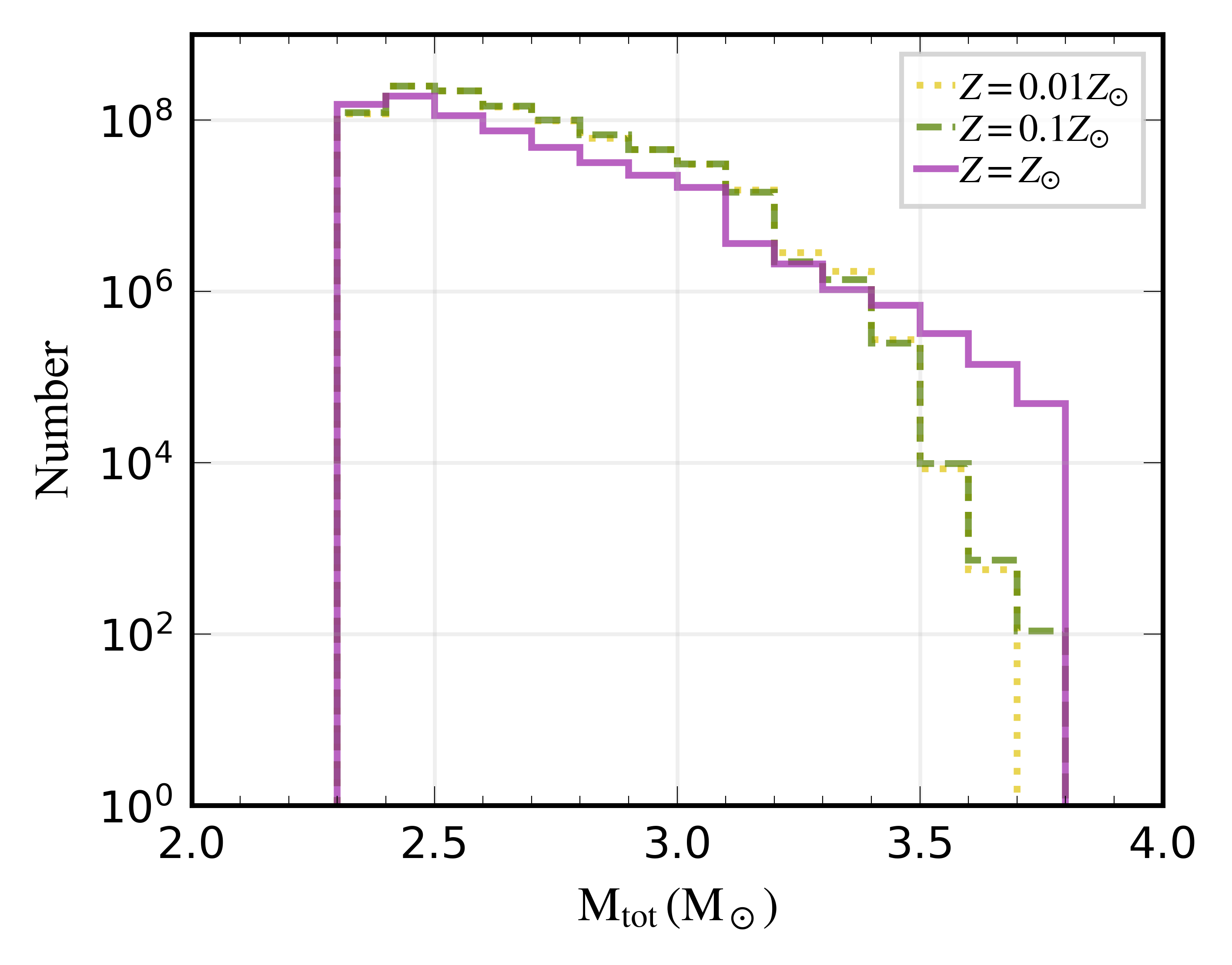}
    \caption{Distribution of the $M_{\rm{tot}}$ from the simulation using remnant mass prescription given by \citet{2012ApJ...749...91F}, and the natal kick velocity of newborn NSs (NS2) from the Hobbs-Maxwellian distribution \citep{Hobbs_2005}. We show the distribution at 
    three different metallicities represented with green, orange, and blue solid lines corresponding to Z = $\mathrm{Z_{\odot}}$, 0.1\,$\mathrm{Z_{\odot}}$, and 0.01\,$\mathrm{Z_{\odot}}$, respectively.}
    \label{fig:fryer_dns_hist}
\end{figure}

\begin{figure}
    \centering
    \includegraphics[width =\columnwidth]{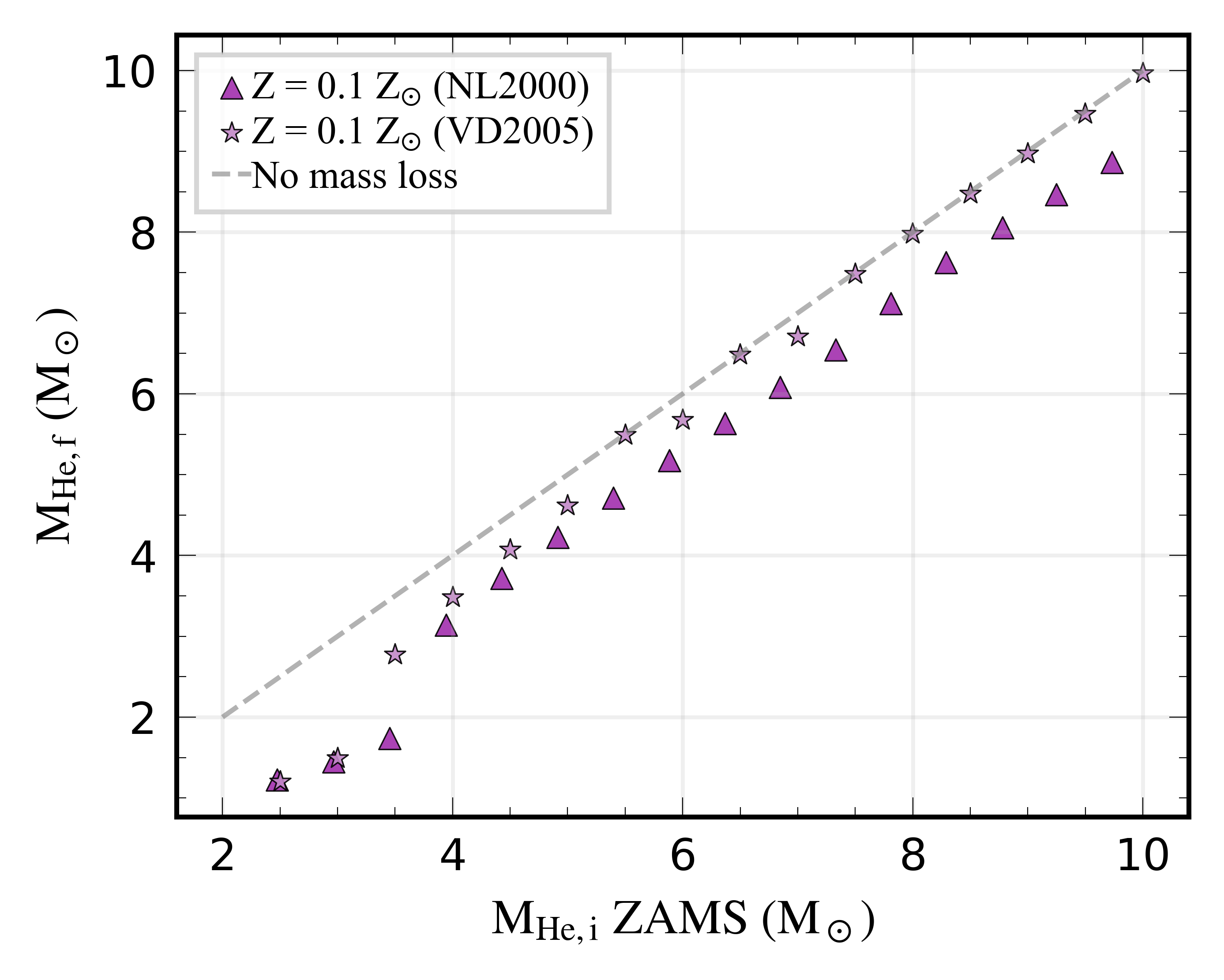}
    \caption{Final helium star mass $M_{\rm{He,f}}$ as a function of initial helium star ZAMS mass $M_{\rm{He,i}}$ for our simulated binary models with an initial orbital period of 0.145 d at 0.1 $Z_\odot$. The grey dashed line corresponds to no mass loss for reference. Both wind mass loss recipes fall below this line, showing wind mass loss via winds from the helium star before the supernova explosion. NL2000 \citep{2000A&A...360..227N} recipe, represented by triangles, shows more mass loss compared to the VD2005 \citep{2005A&A...442..587V} recipe, which introduces metallicity dependence, with the discrepancy appearing towards high-mass HeZAMS stars.}
    \label{fig:NL_VD_Mhe_i_vs_f_Porb0.145d}
\end{figure}

\begin{figure}
    \centering
    \includegraphics[width =\columnwidth]{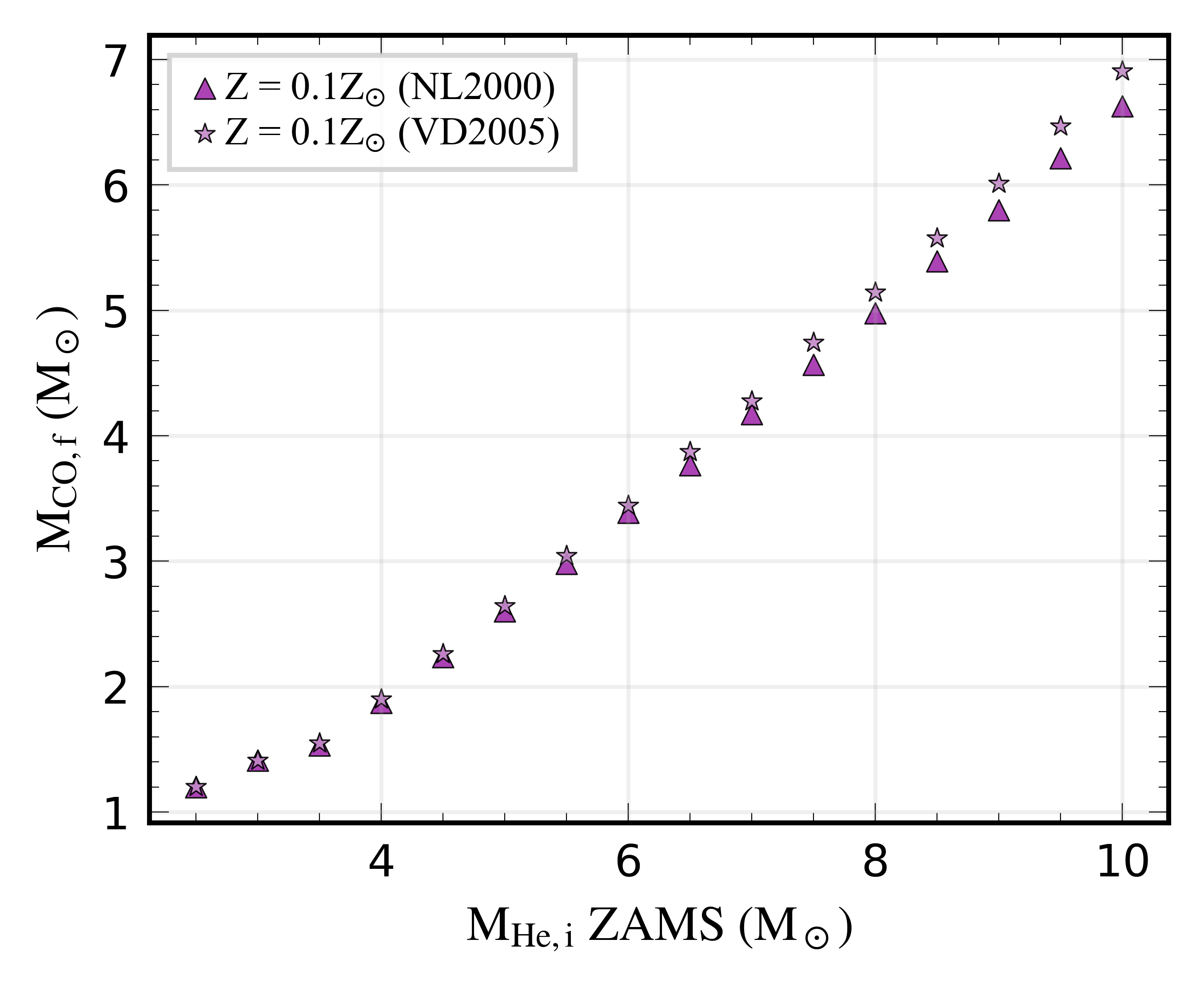}
    \caption{Final carbon-oxygen (CO) core mass ($M_{\rm{CO,f}}$) as a function of initial helium star ZAMS mass ($M_{\rm{He,i}}$) for our simulated models at 0.1 $Z_\odot$ and initial orbital period $P_{\rm{orb}} = 0.145$\,d comparing the NL2000 \citep[][triangles]{2000A&A...360..227N} and VD2005 \citep[][stars]{2005A&A...442..587V} wind mass-loss recipes. The two recipes produce nearly identical $M_{\rm{CO,f}}$ across the full mass range, with only a gradual divergence above $M_{\rm{He,i}}$ $\ge$ 8\,\msun{}, indicating that in this scenario the final CO core mass is largely insensitive to the choice of wind prescription.}
    \label{fig:NL_VD_MCO_i_vs_f_Porb0.145d}
\end{figure}

\begin{figure}
    \centering
    \includegraphics[width =\columnwidth]{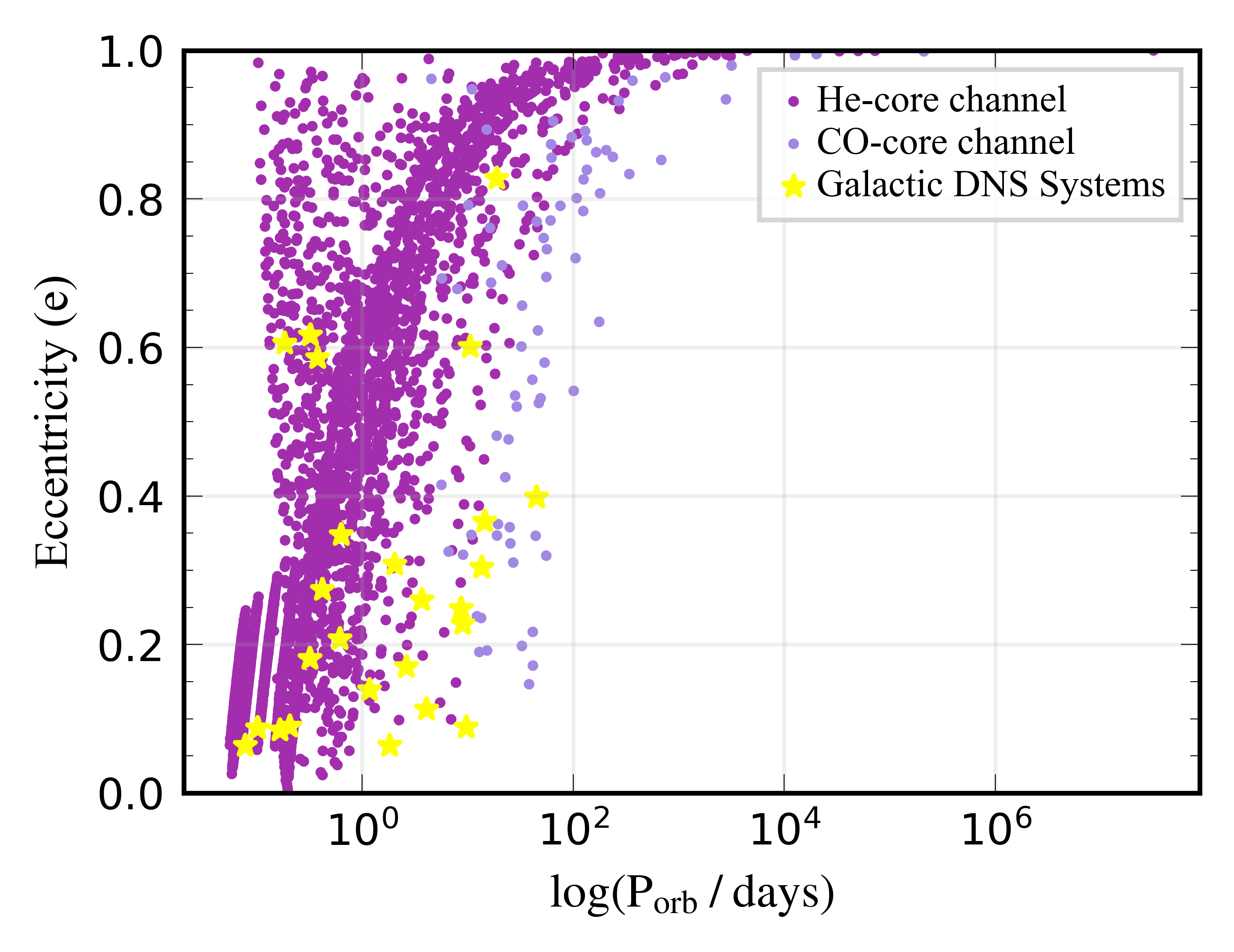}
    \caption{Post-supernova orbital eccentricity (e) as a function of orbital period $P_{\rm{orb}}$ for simulated double neutron stars formed from a He star mass of 3.5\,\msun{} with a first-born NS companion of 1.3\,\msun{} at solar metallicity. The systems originate from two sub-channels introduced by \citet{Chattaraj:2026ApJ}, the helium-core (He-core, dark purple) and carbon-oxygen core (CO core, light purple), compared with observed Galactic DNS systems (yellow stars).}
    \label{fig:he_co_core_bifurcation}
\end{figure}

\newpage


\bsp	
\label{lastpage}
\end{document}